\documentclass[10pt]{article}

\usepackage[utf8]{inputenc}
\usepackage[T1]{fontenc}
\usepackage{lmodern}
\usepackage{microtype}
\usepackage[margin=2cm]{geometry}
\usepackage{graphicx}
\usepackage{amsmath,amssymb,amsthm}
\usepackage[hyphens]{url}
\usepackage{booktabs}
\usepackage{caption}
\usepackage{subcaption}
\usepackage{enumerate}
\usepackage{fancyhdr}
\usepackage{lineno}
\usepackage{xcolor}
\usepackage{authblk}
\usepackage{multicol}
\usepackage{soul}

\usepackage{xcolor} 
\usepackage{graphicx}
\usepackage{booktabs}
\usepackage{tabularx,ragged2e}
\newcolumntype{C}{>{\Centering\arraybackslash}X} 
\usepackage{authblk} 
\usepackage{appendix}

\usepackage[backend=biber, style=nature, autocite=superscript,  doi=false, url=false, isbn=false]{biblatex}
\usepackage[colorlinks=true,linkcolor=blue,citecolor=blue]{hyperref}

\makeatletter
\renewcommand{\@maketitle}{%
  \newpage
  \null
  \vskip 1em%
  \begin{center}%
    \let \footnote \thanks
    {\Large\bfseries \@title \par}%
    \vskip 1em%
    {\normalsize
      \lineskip .5em%
      \begin{tabular}[t]{c}%
        \@author
      \end{tabular}\par}%
    \vskip 1em%
    {\normalsize \@date}%
  \end{center}%
  \par
  \vskip 1em}
\makeatother

\usepackage{titlesec}
\titleformat{\section}
  {\normalfont\large\bfseries}{\thesection}{1em}{}
\titleformat{\subsection}
  {\normalfont\normalsize\bfseries}{\thesubsection}{1em}{}

\renewenvironment{abstract}{%
  \small
  \begin{center}
    \bfseries Abstract
  \end{center}
  \quotation
}{%
  \endquotation
}

\begin{document}

\title{Political attitudes differ but share a common low-dimensional structure across social media and survey data} 

\author[a,b]{Antoine Vendeville}
\author[b,c]{Hiroki Yamashita}
\author[b,a]{Pedro Ramaciotti}

\affil[a]{Sciences Po médialab, Paris, France}
\affil[b]{Complex Systems Institute of Paris Ile-de-France CNRS, Paris, France}
\affil[c]{École des Hautes Études en Sciences Sociales, Paris, France}

\date{}

\maketitle

\begin{abstract}
Does polarization online reflect the state of polarization in society? We study ideological positions and attitudes on several issues in France, a country with documented issue nonalignment. We compare distributions on X/Twitter with a nationally representative sample, focusing on two key properties: ideological polarization and issue alignment. Despite significant issue-wise divergences, positions of both the X population and the nationally representative sample present a similar bi-dimensional structure along two dominant bundles of aligned issues: a Left-Right divide, and a Global-Local divide.
We then study how our results vary when accounting for key structural parameters of the online public sphere: activity, popularity, and visibility.
We find that the dimensionality of attitude distributions shrinks as ideological polarization increases when selecting more active users. The divergence between political attitudes on social media and in survey data is greatly mediated by the combination of activity and popularity of social media users: users benefiting from the most exposure are also the most representative of the general public.
Together, our results shed light on the structural similarities and differences between political attitudes from social media users and the general public.

\end{abstract}

\vspace{1.0cm}

\begin{multicols}{2}

\section{Introduction}

The study of political polarization has sparked widespread interest across academic disciplines in the last decades. Different forms of political polarization have been identified in the literature \autocite{jost2022cognitive}, including ideological and issue polarization, as well as issue alignment. The former refers to the distancing of individuals along ideological scales or issue dimensions \autocite{dimaggio1996have}. Some scholars argue that Left- and Right-wing citizens have grown further apart \autocite{ojer2025charting,abramowitz2008is}, while others point to the inconclusiveness of evidence suggesting increasing political polarization \autocite{fiorina2008political,traber2023group}.
Issue alignment refers to the existence of correlation patterns between attitudes toward different political issues. How do these correlations delineate ideological and partisan groups is of particular interest \autocite{baldassarri2008partisans}. American scholars have hypothesized that the increased sorting of partisans along a single Liberal-Conservative axis is a key factor in the growing resentment between Democrat and Republican voters \autocite{finkel2020political,mason2015disrespectfully}, thus relating issue alignment with affective polarization \autocite{iyengar2019origins}. As attitude distribution collapses onto a single axis, the U.S. political landscape is typically (though not systematically \autocite{ojer2025charting,uscinski2021american}) thought of as a single-dimensional space. 

By contrast, scholarship agrees on the multidimensional nature of political competition and behavior in most European countries \autocite{mccoy2018polarization,bakker2012exploring,stoll2010elite,bornschier2010new}, linked to the multi-party nature of European politics \autocite{lijphart_democracies_1984,taagepera_number_1999}. Bottom-up cleavage theory explains the relation between parties and dimensions by positing that party competition reflects historically determined structures of socio-economic conflict \autocite{lipset_cleavage_1967,bartolini_identity_2007}. To accommodate the diverse interests generated by combinations of social cleavages, a greater number of parties emerges. From a top-down strategic perspective, multi-party systems enable parties to introduce or activate dimensions of competition \autocite{de_vries_when_2012}, thereby disrupting the policy linkages established by mainstream parties and offering alternative issue packages to voters \autocite{rovny_struggle_2012}.

The European political landscape has been classically conceptualized along two dimensions, the cultural and the economic Left-Right dimensions. But in the wake of the twenty-first century, social and economic globalization have changed the meaning of classic cultural and economic dimensions, reflecting new political conflicts between winners and losers of globalization \autocite{kriesi_globalization_2006}. Globalization is particularly epitomized in Europe by the issue of EU integration, being understood simultaneously as a social democratic project and as a challenge to national sovereignty \autocite{marks_past_2000}. Two recent developments have potentially intensified the centrality of this Global-Local divide. First, challenger parties exploited this new political conflict \autocite{de_vries_political_2023}, while mainstream parties took accommodation strategies after the success of far-Right parties \autocite{abou-chadi_causal_2020}. Second, underlying social transformations, notably the decline of traditional class and religious cleavages \autocite{kriesi_west_2008,marcinkiewicz_religious_2022} have opened grounds for the emergence of new lines of division \autocite{bornschier_cleavage_2024}.

Most results about the structure of the political competition at party level do not directly translate to the public.
Research shows that issue or ideology positions of electorates are less structured than those of political elites \autocite{wheatley_reconceptualizing_2021,kurella_unfolding_2025}. 
In addition, part of the debate among the public is increasingly perceived through the lens of social platforms.
Studies have shed light on recurring patterns of ideological (or issue) polarization \autocite{salloum2025anatomy,quelle2025bluesky,loru2025ideology,flamino2023political,falkenberg2022growing,waller2021quantifying,cinelli2020covid} and affective polarization \autocite{falkenberg2024patterns,lerman2024affective,efstratiou2023nonpolar,williams2015} in social media, while issue alignment has only recently started to gather attention. Research has unveiled alignment patterns between positions on Covid, migration and climate change in online debates across Austria, Germany and Finland \autocite{fraxanet2025unpacking,salloum2025anatomy,pournaki2025how,loru2025ideology,chen2021polarization}, uncovering indices for the existence of a Global-Local divide, unaligned with the traditional Left-Right spectrum \autocite{fraxanet2025unpacking,pournaki2025how,chen2021polarization}.

Social media are widely used as political forums, but often misrepresent the plurality of political opinions \autocite{dimartino2025ideological}. Their users are disproportionately young and urban \autocite{mellon2017twitter,barbera2015understanding,filho2015twitter,malik2021population,mislove2021understanding,vaccari2013social}. An additional self-selection process separates active from passive users, as more ideologically extreme people are more inclined to engage in online discussions \autocite{chen2025public,bessi2016,vaccari2016,barbera2015understanding}, and are rewarded with higher social feedback than moderate voices \autocite{garimella2018political}. These observations however do not speak of the effect of social media on the broader public debate \autocite{guess2023reshares,barbera2015how}. 

Because of the importance of social media debates in shaping perceptions on the public sphere, and particularly perceptions on polarization, we conduct a study to compare ideological polarization and issue alignment between a nationally representative panel and a social media panel.
Our study focuses on France, a multi-party, multi-dimensional political system, and on X (previously Twitter), in 2023. This corresponds to a period by which the platform had considerable penetration in the country, marking the end of decade of deep entanglement between elite, journalistic, and public political communication.
Despite subsequent partial abandonment of X by the public and political figures, the period for the data in our study targets a landmark in the history of the entanglement between offline politics and any social platform.

We consider an X panel consisting of almost a million French users and their positions on a variety of political dimensions \autocite{vendeville2026mapping}. These positions were computed using ideology scaling methods calibrated and identified with political survey data \autocite{ramaciotti2022inferring}, and validated through manual and automated stance annotations. We compare these position distributions with those from the European Social Survey (ESS), which contains self-declared positions of a representative sample of the French population \autocite{ESS2023}. We include eight political dimensions of interest, comprising a variety of ideological and issue dimensions relevant to the French political debate.
Finally, we analyze how key structural variables that determine public perception in social media---activity, popularity, visibility---affect polarization, dimensionality, and divergences between online and offline political positions.

Our results show that French X users are less polarized ideologically than the general public. The X panel is strongly skewed towards Right-wing and anti-establishment positions, while the ESS panel exhibits more moderate positions. Despite these significant differences, we find that political positions in both panels are structured around a Left-Right divide (pertaining to issues such as immigration or redistribution of wealth) and a Global-Local divide (pertaining to issues such as EU integration and anti-elitism). This structure is remarkably more salient in X, explaining at least 90\% of the variance in opinions, versus only 50\% for the ESS panel. Finally, we show that more active users are ideologically more polarized, exhibit more aligned positions, and are less representative of the general public. The vast majority of them, however, are not very popular. The most visible users on the platform exhibit opinions that are less polarized ideologically, skewed toward Left-wing and Globalist positions, and overall closer to those of the general public.
\section{Results}

To study the French X population, we leverage a publicly available dataset \autocite{vendeville2026mapping}, collected in February 2023 and pertaining to the X accounts of 883 French Members of Parliament (MPs) and 978,050 of their followers. The users are endowed with spatial positions in a multidimensional political space that reflect their stances on various ideology axes and political issues. These positions take the form of continuous values between 0 and 10, computed via ideology scaling, calibrated and identified with political survey data, and validated against political stances declared by the users in their profile's self-description (see ``Methods''). The dataset also contains indicators of activity (tweeting rate, number of followees) and popularity (number of followers) for each account. We focus specifically on the MPs' followers, henceforth referred to as the ``X panel'', a set of spatially rational individuals \autocite{luskin1990explaining} coherent with the structure of national politics \autocite{ramaciotti2022inferring}.

To compare the X population with the general public, we rely on the 2023 wave of the European Social Survey (ESS) \autocite{ESS2023}. The ESS is a respondent survey that collects socio-economic data, political attitudes and beliefs from nationally representative panels across European countries. Political attitudes are encoded along Likert-type scales of various lengths. After filtering out missing values, the French panel contains 1,417 respondents in 2023, henceforth referred to as the ``ESS panel''. In accordance with the design of the ESS \autocite{Kaminska2020guide}, we take into account individual weights, stratum and primary sampling units in all of the computations (see ``Methods'').

For the purpose of our study, we select eight political dimensions that are comparable across the ESS and the X dataset: Left-Right, Anti-elitism, EU integration, Immigration, Redistribution of wealth, Social liberalism, Environment, and Direct democracy (support or opposition to referendums). These dimensions are selected based on three criteria: their relevance to the French political debate, the possibility to find proxies of positions in the two sources of data, and the validity of the estimated positions of X users along these dimensions. We indicate the political dimensions and their ordering in both sources of data in \autoref{tab:dimensions}. Details about the scales and precise survey questions corresponding to the chosen dimensions are given in Supplementary Section 1. 

\begin{table*}
\centering
\caption{\textbf{Political dimensions of interest.} Description and ordering of the political dimensions considered. The ordering describes the qualitative variation of positions from the lower to the upper bound of the axes.} 
\begin{tabular}{lcr}
\toprule
\textbf{Dimension} & \textbf{Ordering} & \textbf{Description (ESS)} \\
\midrule
Left-Right & Left $\longleftrightarrow$ Right & Placement on Left-Right scale \\
Anti-elitism & Low $\longleftrightarrow$ High & Trust in politicians \\
EU integration & Pro $\longleftrightarrow$ Anti & European unification go further or gone too far \\
Immigration & Pro $\longleftrightarrow$ Anti & Immigrants make country worse or better place to live \\
Redistribution & Pro $\longleftrightarrow$ Anti & Government should reduce differences in income levels \\
Direct democracy & Anti $\longleftrightarrow$ Pro & Political system allows people to have a say in what government does \\
Environment & Pro $\longleftrightarrow$ Anti & Important to care for nature and environment \\
Social liberalism & Pro $\longleftrightarrow$ Anti & Gays and lesbians free to live life as they wish \\
\bottomrule
\end{tabular}
\label{tab:dimensions}
\end{table*}

\subsection{Dimension-wise comparison of online and offline positions}

\begin{figure*}
    \centering
    \includegraphics[width=\textwidth]{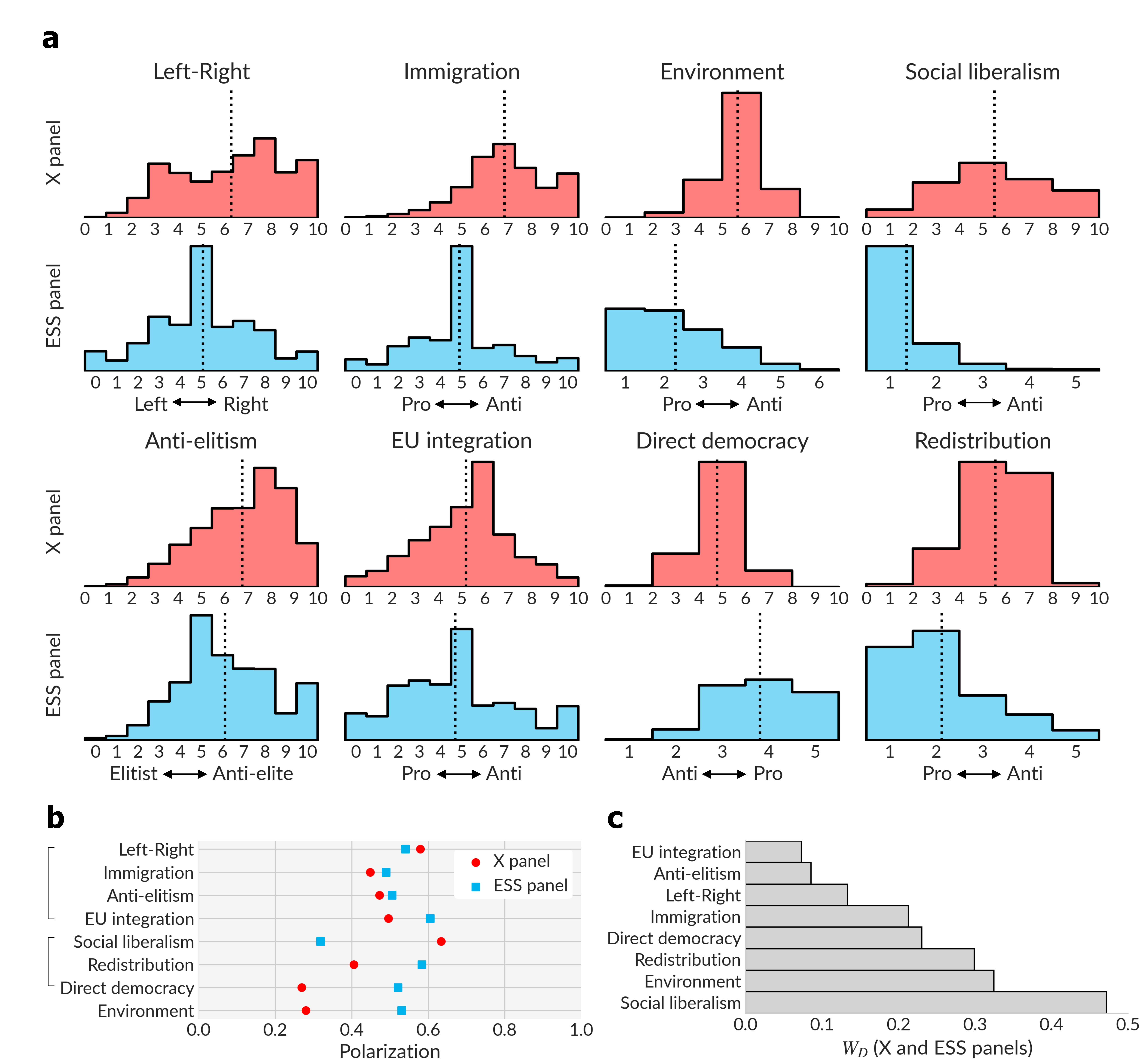}
    \caption{\textbf{Distributions of positions inactive the X panel and the ESS panel.} \textbf{(a)} Histograms of attitude and ideology position distributions. Dotted vertical lines indicate the mean. Distributions for the X panel are sorted into bins to facilitate comparison with the ESS panel. \textbf{(b)} Polarization values across dimensions. Brackets indicate groups of dimensions that share the same support, and within which polarization values can be compared. \textbf{(c)} Normalized Wasserstein distance between X and ESS distributions, sorted from lowest (top) to highest (bottom).}
    \label{fig:epo_vs_ess}
\end{figure*}

We start by comparing dimension-wise distributions of political positions in the X and ESS panels. To facilitate comparison, we discretize the X distributions along the same range as ESS, by sorting them into the adequate number of equal-length bins. Results are shown in \autoref{fig:epo_vs_ess}a.

The distributions are either equally spread along the spectrum with a high concentration of moderate users (e.g., Left-Right in ESS), or skewed on one side (e.g., Immigration in X). X users are generally right-wing, anti-elitist, anti-EU and anti-immigration, with more moderate stances on other dimensions. The dichotomy between clearly skewed positions on certain dimensions and moderate ones on others is reversed for ESS respondents. They are, on average, rather centrist, with moderate views on immigration, EU integration, and slight Anti-elitism (the overabundance of middle-point answers may be partially due to documented respondent bias \autocite{baka2012neither}). Where X users exhibit moderate stances, ESS respondents demonstrate remarkable consensus, clearly favorable to environmental-friendly policies, social liberalism, direct democracy and redistribution.

No distribution exhibits a clear bimodal shape in either datasets, challenging the idea of highly polarized political positions. We evaluate polarization along each dimension, using a non-parametric entropy-based measure proposed by Bao and Gill \autocite{Bao_Gill_2024}. Values range between 0, for a distribution with a single outcome of positive probability, and 1, for a distribution with two outcomes of equal probability. Results are shown in \autoref{fig:epo_vs_ess}b. Polarization values are only comparable across dimensions that share the same support, indicated by brackets. However, we can proceed to dimension-wise comparisons between the two panels. We find higher levels of polarization among ESS respondents on all dimensions except Left-Right and social liberalism. Social liberalism, in particular, appears much more polarized online than offline. Looking at the distribution of the two panels along this dimension reveals a mostly consensual socially liberal view among the general public, opposed to much more spread positions among X users.

To quantify the representativeness of the X panel, we compute the Wasserstein distance ($W_D$) between the two panels on each dimension. The Wasserstein distance, also known as the earth-mover's distance, measures the distance between two distributions by the number of unit displacements that are required to transform one distribution into the other (see ``Methods''). To ensure comparability across dimensions, we rescale $W_D$ by its maximum possible value so that it ranges from 0 (identical distributions) to 1 (the masses of the two distributions are concentrated at opposite sides of the value range). X and ESS panelists are furthest away on Social liberalism ($W_D>0.4$), echoing our previous observation. They are remarkably close on EU integration and anti-elitism ($W_D<0.1$), and somewhat close on Left-Right ($W_D<0.15$). X users are thus most representative of the general public along these three dimensions.

The low levels of polarization we observe do not concord with similar analyses, which most often uncover strong bimodality in ideological positions \autocite{falkenberg2024patterns,loru2025ideology}. This might be due to the fact that existing analyses are based on retweet interactions, which are more homophilic than follow links \autocite{vendeville2025voter}. Moreover, retweet data might exhibit biases due to the unequal levels of activity across users \autocite{dibona2024sampled} (Supplementary Figure 7). On the other hand, most users follow an ideologically diverse array of accounts \autocite{barbera2015how}, and our collection of follow links is exhaustive (see ``Methods'').

\subsection{Alignment between political dimensions}

Having unveiled important dimension-wise divergences between the online and offline populations, we now focus on the alignment between dimensions.

\begin{figure*}[t]
    \centering
    \includegraphics[width=\textwidth]{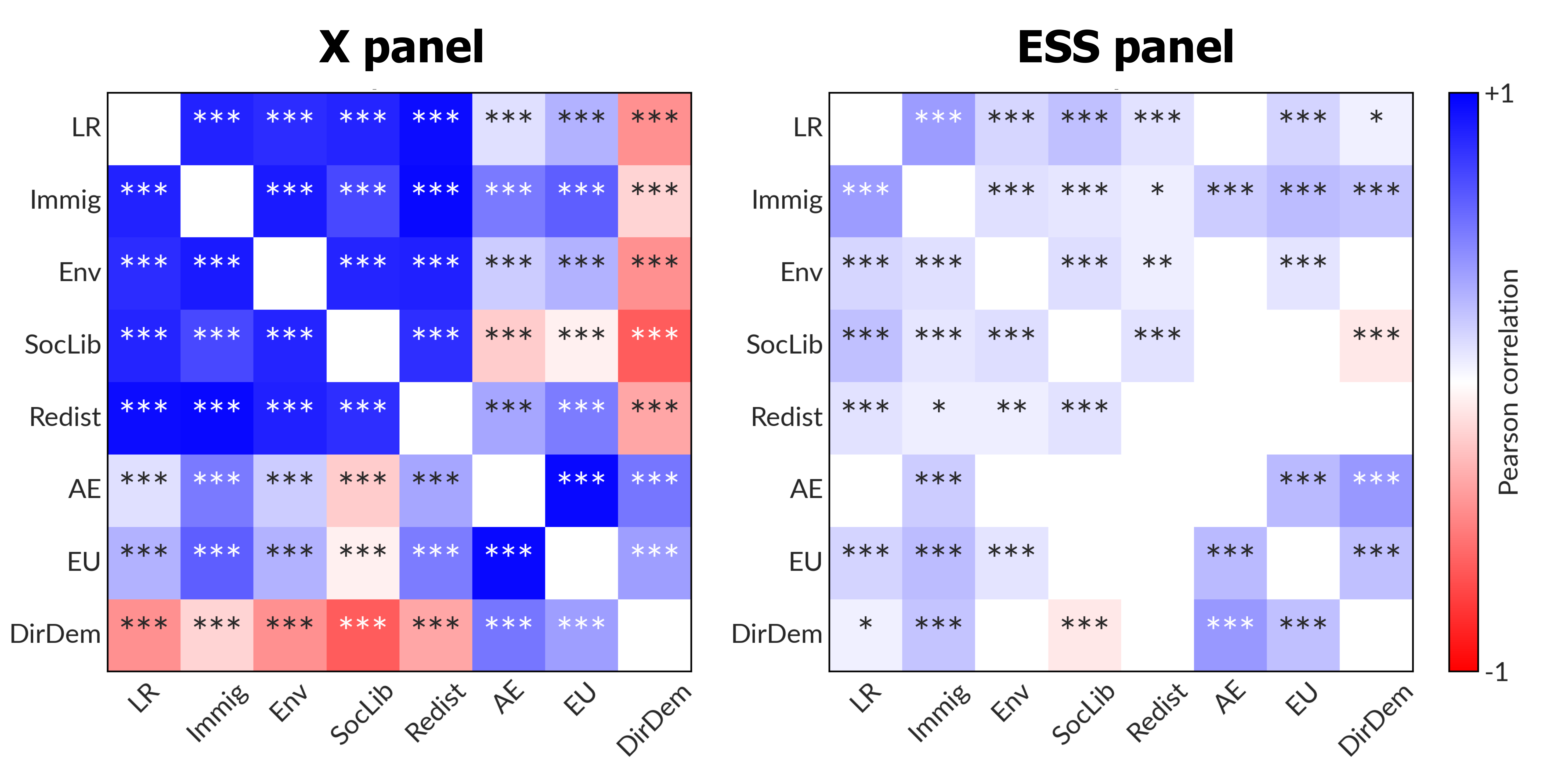}
    \caption{\textbf{Pearson correlations between dimensions.} Significance levels *: $p<.05$, **: $p<.01$, ***: $p<.001$. Immig: immigration, SocLib: Social liberalism, Env: environmental policies, Redist: redistribution, DirDem: Direct democracy. Percent of significant correlations: 100\% (X), 71\% (ESS). Average value of significant off-diagonal correlations: 0.56 (X), 0.18 (ESS). Strongest off-diagonal correlation: $0.97$ (X, Anti-elitism and EU integration), $0.41$ (ESS, Anti-elitism and Direct democracy). Each matrix differs from the identity (Bartlett's sphericity test: $\chi_2>300, p<0.001$). Non-significant correlations and diagonal cells are colored in white.}
    \label{fig:correlations}
\end{figure*}

\autoref{fig:correlations} shows the Pearson correlations between dimensions in each panel. Fewer dimension pairs exhibit strong correlations in the ESS panel (average absolute value of off-diagonal significant correlations: 0.56 for X vs.\ 0.18 for ESS). All correlations outside the diagonal are significant at the level $p<0.001$ level for the X panel. The positions of individuals in the X panel are thus much more constrained.

A common structure emerges from the two panels. The dimensions can be sorted into two groups that present high in-group correlations (in absolute value). The first group pertains to the Left-Right divide, combining cultural and economical issues: Immigration, Environment, Social liberalism, Redistribution, and the corresponding Left-Right ideological axis. The second group combines dimensions related to populism (Anti-elitism, Direct democracy) with the EU integration issue. This axis opposes elitist, pro-globalization panelists to anti-elite, protectionist ones. We call this second group the ``Global-Local'' divide by analogy with previous research on political dimensional identification on X \autocite{ramaciotti2021unfolding}. 

While a similar macroscopical structure of constraints between issue attitudes exists both among ESS respondents and X users, this structure appears to be more salient and lower-dimensional in the online space, due to the overall stronger correlations. The average in-group absolute correlation is about 0.81 for the X panel (respectively out-group, 0.35), with a ratio between in-group and out-group of about 2.32, against 2.09 for the ESS panel (0.19 in-group, 0.09 out-group). Remarkable cross-group correlations for the ESS panel include Immigration and Global-Local attitudes, as well as EU integration and Left-Right ideology.
 
\subsection{Low-dimensional structure}

The analysis of correlations suggest that political attitude distributions are structured in a similar manner across the two panels: a Left-Right and Global-Local divide, with stronger alignment among X than ESS panelists. To unveil the underlying structure of attitude distribution, we seek a lower-dimensional representation of each panel. 

We perform on each panel a Principal Component Analysis \autocite{Jolliffe2002Principal} (PCA), finding a small number of orthogonal axes in the high-dimensional space of available dimensions, onto which the data can be projected. These axes are called the Principal Components and are sorted by descending order of explained variance. 
Additionally, we compute the Effective Dimensionality \autocite{Koedam_Binding_Steenbergen_2025} (ED) of each panel, a measure of the number of dimensions necessary to describe the structure of multidimensional political spaces, closely related to PCA (see ``Methods''). $ED=1$ means that all political dimensions are perfectly correlated and the positions of the panelists can be expressed on a single line without loss of information (e.g., liberal-conservative); $ED=2$ means that political positions can be expressed on a bidimensional plane (e.g., social issues and economical issues), and so on. If the dimensions are all uncorrelated then $ED=8$, and the political positions cannot be expressed in a lower-dimensional space without important loss of information.

\begin{figure*}[t]
    \centering
    \includegraphics[width=\textwidth]{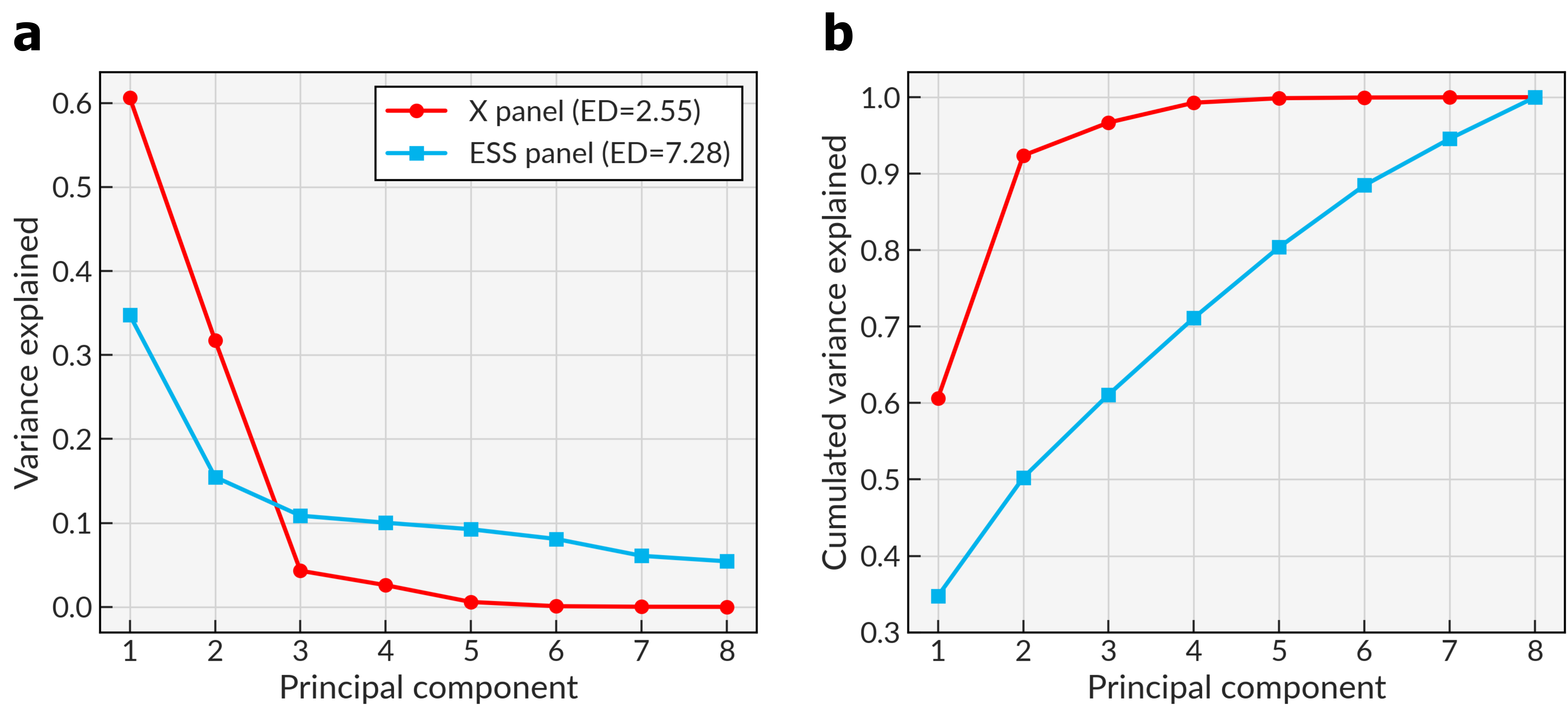}
    \caption{\textbf{Variance explained by the principal components.} We show explained \textbf{(a)} and cumulated explained variance \textbf{(b)} of populations on the European Social Survey (ESS) and X (previously Twitter). ED stands for Effective Dimensionality.}
    \label{fig:expl_var}
\end{figure*}

\autoref{fig:expl_var} shows the variance and the cumulated variance explained by each Principal Component, alongside with the effective dimensionality of each panel. The X panel exhibits an almost bi-dimensional structure, with the first two PCA components (hereafter, PC1 and PC2) explaining more than 90\% of the variance together and an effective dimensionality of 2.55. In contrast, ESS respondents require at least seven components to explain 90\% of the variance, and two to explain 50\%.

These observations echo our previous findings: the distributions of attitudes of X users are well expressed in a two-dimensional space, meaning an ideology sorting along two dividing lines, that we expect to be the Left-Right and Global-Local divides hypothesized after inspection of the correlation matrices. Although from the previous inspection of the correlations matrices, the same two divides appear to be relevant for the ESS panel, much less variance is explained by the first two PCA components in these cases, owing to the overall more moderate positions of individuals in the ESS panel.

\begin{figure*}
    \centering
    \includegraphics[width=\textwidth]{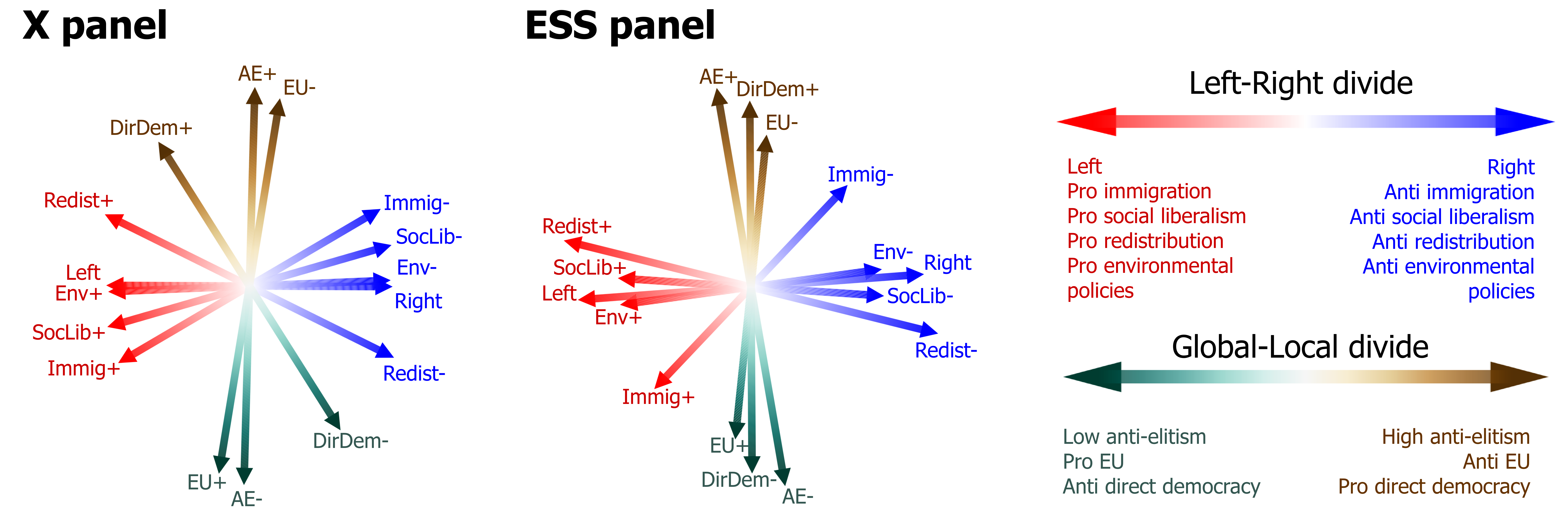}
    \caption{\textbf{Loadings of political dimensions along principal components.} Each arrow shows the loading of a political dimension along PC1 (x-axis) and PC2 (y-axis). Arrow length is proportional to the absolute value of the loading. Red-blue arrows correspond to dimensions related to the Left-Right attitudes, and green-brown arrows correspond to dimensions related to Global-Local attitudes Plus (+) signs are short for ``pro'', and minus (-) signs for ``anti''.}
    \label{fig:shining_suns}
\end{figure*}

For a more systematic characterisation of the low-dimensional space unveiled by the PCA, we inspect the composition of PC1 and PC2, rotated to enhance interpretability (see ``Methods''). \autoref{fig:shining_suns} displays the loadings of the political dimensions in the bidimensional space spanned by PC1 and PC2: the arrows show how the political positions of the panelists are ordered in the (PC1, PC2) plane, and the arrow lengths indicate how well the dimensions are summarized by the PCs. For instance, the Left-Right and the Environment dimensions are almost perfectly aligned with PC1 for the X panel, meaning that the positions of X panelists along these two dimensions are well summarized by a single axis in a low-dimensional space.

In this low-dimensional space, political dimensions are clearly grouped in two clusters, corresponding to the hypothesized Left-Right and Global-Local divides (blue-red and green-brown arrows, respectively). A few bridging dimensions appear in the X and ESS panels. Immigration in particular overlaps the two PCs in ESS, showing the particular relevance of this dimension along the two main dividing lines. For the X panel, Redistribution and Direct democracy also bear this role. The lower effective dimensionality of the X panel is reflected by the greater length of the arrows, showing that each dimension loads heavily on PC1 or PC2. This differs for the ESS panel, for instance with Social liberalism. In fact, the third PC is composed almost exclusively of Social liberalism, against Direct democracy for the X panel (Supplementary Figure 1). These dimensions are thus less aligned with others, not falling as neatly onto the (Left-Right, Global-Local) structure.

\begin{figure*}
     \centering
     \includegraphics[width=\textwidth]{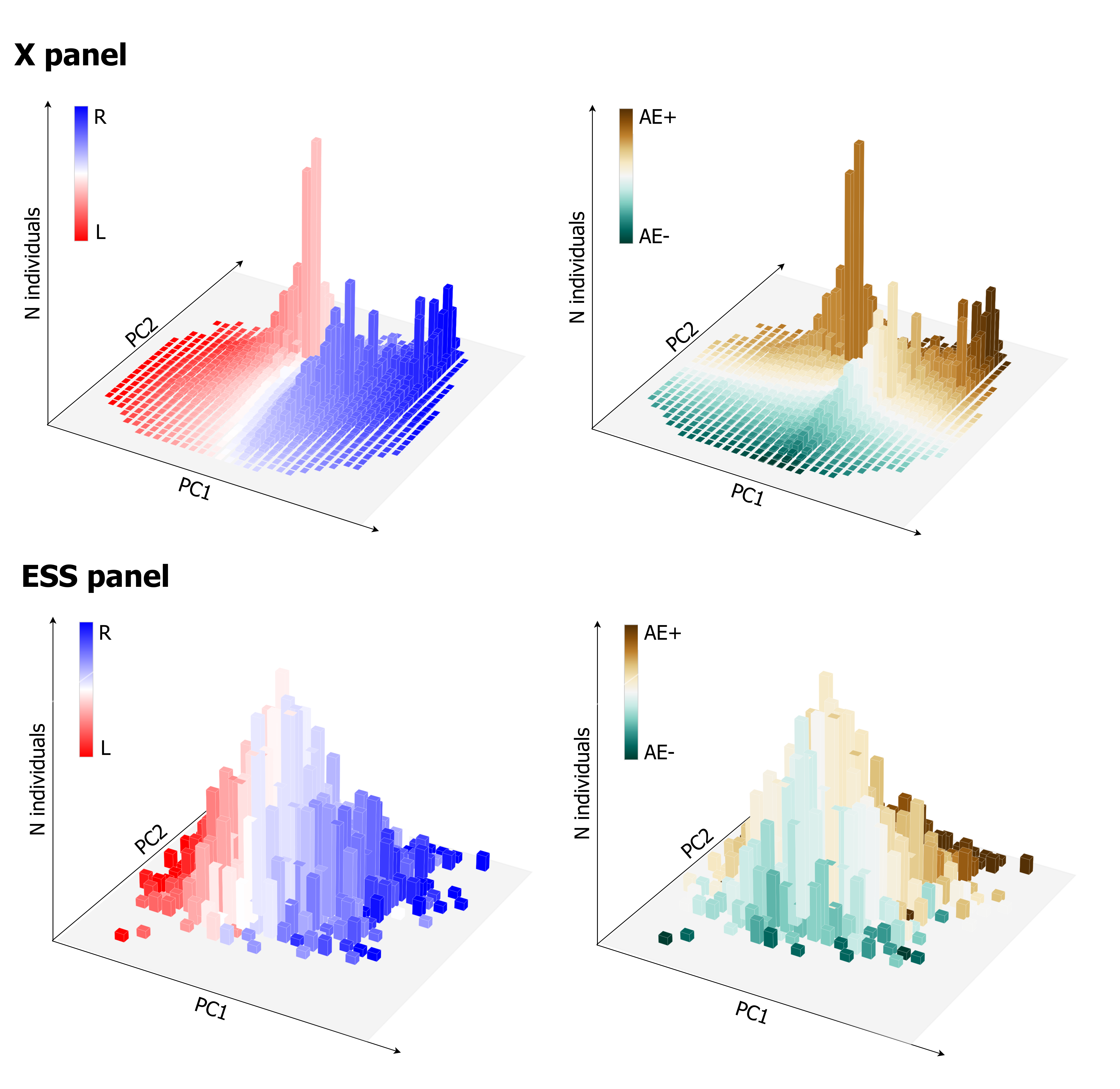}
    \caption{\textbf{Distribution of panelists in the low-dimensional space.} The (PC1,PC2) plane is divided into a grid. Vertical bars indicate the number of individuals in each cell. Bar colors map the average position of individuals on the Left-Right and Anti-elitism dimensions. \textbf{(a,b)} X panel, grid size $30\times 30$. \textbf{(c,d)} ESS panel, grid size $20\times 20$.}
    \label{fig:3D}
\end{figure*}

We inspect in \autoref{fig:3D} the distribution of X and ESS panelists in the low-dimensional space. Bar heights indicate the density of individuals, while color gradients map average values on the Left-Right (left panel) and Anti-elitism (right panel) dimensions. ESS panelists are distributed in a Gaussian manner, while the high alignment of positions sorts the X panelists into three clearly identifiable poles: an anti-elite Left pole, a Center-Right pole with low or moderate anti-elite stances, and a highly anti-elite Far-Right pole. This result is driven by the more structured online political space, and stands in contrast to the lower polarization of X panelists on individual dimensions.

\subsection{Variations of political positions with activity and popularity}

We now investigate how ideological polarization, issue alignment and distance from the general public vary across the different levels of activity and popularity among X users. We sort X panelists into bins according to their tweeting rate (activity), or number of followers (popularity). The number of panelists in each bin fluctuates in magnitude between the orders of $10^3$ and $10^6$ (Supplementary Figure 7). 

\begin{figure*}[t]
    \centering
    \includegraphics[width=\textwidth]{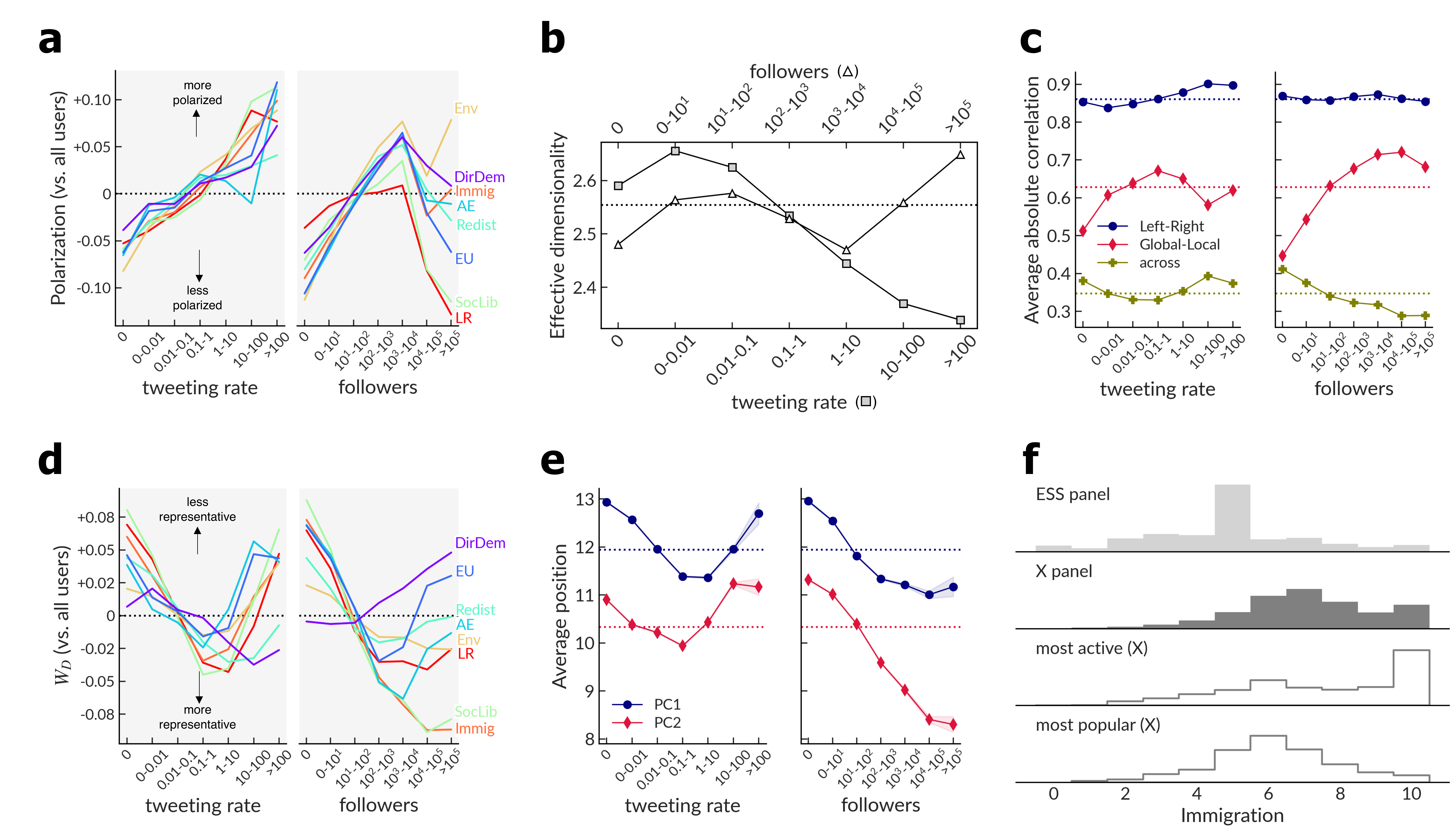}
    \caption{\textbf{Variation of polarization, representativeness and dimensionality of X panelists with activity and popularity.} Users are binned according to their activity (tweeting rate), or their popularity (number of followers). Dotted lines indicate values computed for the whole panel. \textbf{(a)} Polarization along the different dimensions: in-bin value minus overall value. \textbf{(b)} Effective dimensionality. Grey squares denote users in the same activity bin, white triangles users in the same popularity bin. \textbf{(c)}  Average absolute correlation between dimensions within the Left-Right divide (blue circles), within the Global-Local divide (red diamonds), and across the Left-Right and Global-Local divide (green crosses). \textbf{(d)} Wasserstein distance between positions of X users and ESS respondents: in-bin value minus overall value. \textbf{(e)} Average position of users along PC1 (blue circles) and PC2 (red diamonds), with 95\% confidence intervals. \textbf{(f)} Distribution of positions on immigration for the ESS panel, the X panel, the most active X users (tweeting rate $>100$), and the most popular X users (followers $>10^6$).}
    \label{fig:actipop}
\end{figure*}

Sorting users by activity or popularity yields two very different pictures, as shown in \autoref{fig:actipop}. Activity appears to correlate with higher ideological polarization and issue alignment, and popularity with higher representativeness of the positions of the general public.

The more active users are, the more polarized they are: ideological and issue polarization increase with tweeting rate for all political dimensions considered (panel \textbf{a}), albeit to a small extent (maximum increase: 0.1 on a 1-length scale). Dimensionality exhibits an almost monotonous inverse relation with tweeting rate (panel \textbf{b}), dropping to 2.34 for the most active user group. By contrast, the least active users exhibit slightly higher dimensionality than the average. This downward trend in dimensionality might be explained by the reinforcement of correlations between dimensions within the Left-Right divide (panel \textbf{c}), suggesting that this single divide can summarize better than before individual positions.

The picture is more contrasted when users are sorted according to popularity. Both the least and the most followed users exhibit lower polarization than the average, with only Environment displaying a sustained increase of polarization with popularity (panel \textbf{a}). The most polarized users are found in the $[10^3,10^4]$ followers bracket. The variations of effective dimensionality with popularity do not exhibit a particular trend, although we notice that the most popular users exhibit high dimensionality (panel \textbf{b}). This might be due to reinforcement of correlations within the Global-Local divide (panel \textbf{c}), alongside a diminishing of correlations between the Left-Right and Global-Local divides: these two observations point to a decrease of the summarizing power of the Left-Right divide.

Regarding representativeness of the general public, the picture is most clear when sorting users by popularity. Popularity appears highly correlated with representativeness, as the most popular users are also the closest to ESS panelists (panel \textbf{d}). The only exceptions are Direct democracy, and to a lesser extent EU integration, which display the opposite behavior. These results may stem from the downward trend in the position of users along PC1 and PC2 as popularity increases (panel \textbf{e}): the more popular the user, the more Left-wing and Globalist they are, pushing them closer to the ESS panel.

The least and the most active X users on the other hand, form the two groups that are the least representative of ESS respondents along most dimensions (panel \textbf{d}). These users are mostly Right-wing Localists (high values along PC1 and PC2, panel \textbf{e}). Users with a tweeting rate between 0.01 and 10 tweets per day are the only ones to be closer to ESS respondents than the average.

As an illustration, we show in panel \textbf{f} the distribution of positions on immigration for the ESS panel, the X panel, the most active and the most popular X users. The most active users are much more anti immigration, while the most popular are much more pro immigration, thus bringing them closer to the general public described by the distribution of ESS respondents.

\subsection{Visible opinions in the online sphere}

This disparity between the political positions of the most active and the most popular users begs the question: How does X's political sphere appears to the average user? We define the visibility of an X user as $tweeting\ rate \times followers$. This quantity is consistent with the number of times that the users' tweets were viewed, a metric that we have access to for a subsample of users in January and February 2023 (Pearson correlation 0.92, Supplementary Section 4.4). We weight users by their visibility and measure again the polarization of this population. Doing so, we draw a clearer picture of which users benefit from the most exposure on X---putting aside algorithmic effects. Under our visibility metrics and its underlying assumption, users who are not followed or who do not tweet are essentially invisible, while content posted by very active or very popular users is much more likely to appear on the feeds.

\begin{figure*}
    \centering
    \includegraphics[width=.9\textwidth]{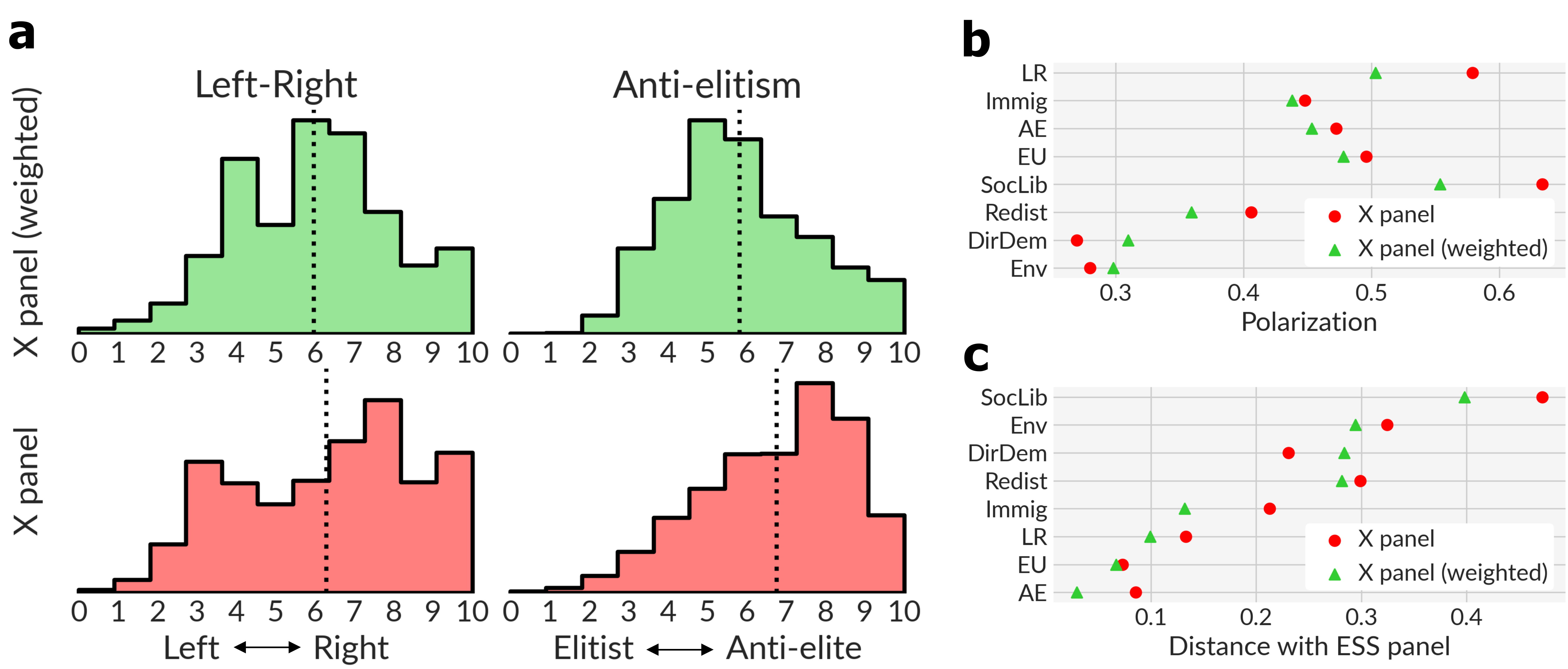}
    \caption{\textbf{Visibility of opinions on X.} We compare political positions between the X panel (red), and the X panel with users weighted by their visibility (green). \textbf{(a)} Distributions of positions for Left-Right and Anti-elitism. \textbf{(b)} Polarization along dimensions. \textbf{(c)} Wasserstein distance with the ESS panel.}
    \label{fig:X_weighted}
\end{figure*}

When X panelists are weighted by their visibility, most of the distributions of political positions are less skewed and more moderate (\autoref{fig:X_weighted}, panel \textbf{a}). In consequence, opinion distributions appear less polarized, and closer to the distribution of ESS panelists (panels \textbf{b,c}). Therefore, the landscape of perceivable opinions of X users is closer to the opinions of the general French population. Regarding issue alignment and dimensionality however, there is no significant change when weighting the users (Supplementary Figure 9).

\section{Discussion}

This work assesses similarities and differences between online political spheres and broader public opinion, focusing on ideological polarization and issue alignment. We studied the distributions of political positions of two populations: a population of users in the French X sphere, for which opinions were derived via ideology scaling methods and calibrated using political surveys data \autocite{vendeville2026mapping}, and a nationally representative panel of French citizens with self-declared attitude and ideology positions, extracted from the European Social Survey \autocite{ESS2023}. 
While the question of the representativeness of the X population is challenging because it naturally lacks referential populations, the panel we use represents users that are directly connected to referential political actors (Members of Parliament in France), and constitutes a population that is spatially coherent.
We studied eight relevant political dimensions: Left-Right, Immigration, Environment, Social liberalism, Redistribution of wealth, Anti-elitism, EU integration, and Direct democracy. 

We found the general public to be more polarized than X users engaged in the political sphere along almost all dimensions. X users are, however, consistently skewed toward Right-wing, populist positions, while the general public exhibits more centrist positions. Regarding issue alignment, X users exhibited very constrained positions, as shown by the strong and highly significant correlations between political dimensions, while correlation patterns for the general public were similar but less significant. These correlation patterns appear to delineate two groups of political dimensions, with high in-group and low out-group correlations. In consequence, both the online and offline political spheres were amenable to a lower-dimensional representation.%

By applying a standard dimensionality reduction procedure to our data, we unveiled a bi-dimensional representation for each population. This representation is structured around a typical Left-Right divide, pertaining to social and economical issues such as immigration and redistribution of wealth, and a Global-Local divide, opposing protectionist, populist attitudes to globalist, elitist attitudes. As a consequence of the higher and more significant correlations, this representation retained more than 90\% of the variation in the original X data, against just 50\% for the general public. While the general public is distributed in Gaussian-like manner in this bi-dimensional space, X users are concentrated around three poles: a Left-wing localist pole, a center-right pole, and a Far-Right localist pole. Our results are robust to an alternative choice of dimensionality reduction method (Supplementary Figure 4).

Finally, we analyzed how these results vary with the activity and popularity of X users, measured by the tweeting rate and number of followers. By selecting more active users, the resulting polarization of distributions increases and the dimensionality of their political opinion space shrinks, indicative of increased issue alignment. Selecting more popular users on the other hand, results in an enhanced representativeness with regards to the general public, with more Left-wing, Globalist positions in the low-dimensional space. To uncover the opinion landscape to which X users are exposed, we reiterated our previous analyses while weighting our sample by their estimated visibility on the platform (tweeting rate multiplied by number of followers). The landscape that emerges appears generally less polarized and closer to the positions of the general public, but we observed no significant changes in terms of dimensionality and issue alignment (Supplementary Figure 9).

Our findings must be contextualized within the political situation in France. At the time of data collection in 2023, a tri-partite structure opposed the governing coalition around Emmanuel Macron (Center-Right), the Right party Rassemblement National (RN, National Gathering), and the Left coalition led by the Radical Left party La France Insoumise (LFI, Unbowed France). The divide between the governing coalition and the two oppositions (LFI and RN, both displaying anti-elitist positions according to expert surveys on party positions) was defined by attitudes toward globalization \autocite{gougou_consolidation_2024}, while LFI and RN were strongly opposed to one another on cultural issues. As such, the positions of these three poles cannot be reduced to a single Left-Right axis. 

Our findings offer two interrelated, crucial insights pertaining to online political debates. First, they restate the importance of considering multidimensional attitude and ideology distributions in the study of online political debates. We show the insufficiency of a single-dimensional representation of political positions of online users, which runs the risk of disregarding as much as 40\% of the variation in opinions (\autoref{fig:expl_var}). A lot of research in online debates has focused on single-dimensional political frames (Liberal-Conservative or Left-Right scales), for two reasons: the difficulty of obtaining fine-grained political positions for large online populations, and the prevalence of U.S.-focused studies. More general national settings, including most European ones, require multidimensional scales to analyze party competition \autocite{mccoy2018polarization,bakker2012exploring,dolezal2013structure,bornschier2010new} and public opinion \autocite{wheatley_reconceptualizing_2021,kurella_unfolding_2025}. As most online studies focus on ideological and affective polarization, our methodology comes as a tool for the study of issue alignment. In particular, the method for the inference of multidimensional opinions from interaction data, described in a previous publication \autocite{vendeville2026mapping}, is easily applicable to vast corpora of online data. 

The second insight of our study pertains to the representativeness of actors engaging in the online public debate when compared to the general public. Many studies have unveiled significant socio-demographically biased selection in online population compared to the general public \autocite{mellon2017twitter,barbera2015understanding,filho2015twitter,malik2021population,mislove2021understanding, vaccari2013social}, with additional selection separating active and inactive users \autocite{chen2025public,bessi2016,vaccari2016,barbera2015understanding,garimella2018political}. The last part of our results demonstrates that, despite significant political skew among X users in our panel, in particular highly active accounts, the most visible accounts---defined as both active and popular accounts---shape the contour of a political landscape that appears remarkably close to the general public's opinion. Further studies shall investigate this complex interplay between activity, popularity and visibility to better understand mechanisms of exposure in online social media.

Our study does not come without limitations. The first concerns the comparability between the X dataset and the ESS. The political positions of X users are computed on the basis of interactions with members of parliament on the platform. As such, these positions are derived from observations of the political \emph{behavior} of the users. By contrast, political opinions of ESS respondents are \emph{self-declared}. This limitation is attenuated by the fact that the positions of the X users were consistent with self-declared opinions, retrieved on the users' online profiles when available \autocite{vendeville2026mapping}. The second limitation of our study pertains to our analysis of the visibility of the X users. Visibility is not only dependent on activity and popularity, it is also highly dependent on algorithmic effects. The study of these effects is an important field of study by itself, rife with in-depth analyses of X \autocite{piccardi2025reranking,bouchaud2023crowdsourced,huszar2022algorithmic}. We argue that our choice of not considering algorithmic effects is a feature rather than a shortcoming. We choose specifically to focus on behavioral indices, to better understand the correlations that exist between online behavior and political positions. 
Our visibility index is directly related to the potential number of views if recommendation was chronological, which is a common baseline against which algorithmic recommenders are often compared.
Integrating algorithmic effects would considerably extend the scope of the article, and thus would be better suited as a separate study.

We hope that our results and observations will encourage future studies. They call for more research dedicated to the study of dimensionality of opinions in social media, and for more evaluations of the representativeness of the general population in online spaces. Future research shall strive to develop similar analyses for other countries. As the pool of relevant political issues and ideological axes may vary, it is particularly important to extend these considerations beyond a single country. Whenever possible, such studies shall also strive to assess how these results may vary over time. Finally, more platforms should be considered, especially in the light of the recent perspective on ``echo platforms'' \autocite{dimartino2025ideological}.
\section{Methods}

\subsection{X dataset} \label{subsubseq:X_data}
The procedure employed to build the X dataset is detailed in the corresponding article \autocite{vendeville2026mapping}. We first collected metadata of the X profiles of all French MPs present on the platform in February 2023, and of their followers. After filtering out accounts without sufficient connections, we built a bipartite network linking MPs and their followers. There are 883 MPs and 978K followers.

Using the ideology scaling method originally proposed by Barberá \autocite{barbera2015birds}, MPs and their followers were embedded in a multidimensional latent ideological space. Distances in this space reflect ideological proximity between the users: it has been shown in several independent studies that the positions along the first dimension of this space are often strongly correlated with Left-Right ideology \autocite{barbera2015birds,bond2015quantifying}. To go beyond the Left-Right axis, and derive individual positions along several political dimensions, we leveraged data from the 2019 and 2023 waves of the Chapel Hill Expert Survey (CHES) \autocite{ches2019,ches2023}. The CHES contains political positions of major parties in European countries, including France, in the form of floating numbers between 0 and 10, along several key ideological axes and political issues, e.g., Left-Right, Anti-elitism, Immigration, EU integration.

Data from the CHES can thus be seen as an embedding of the political parties in a multidimensional hypercube $[0,10]^d$. We established a mapping between this space and our latent space obtained by ideology scaling: we first computed proxies of party positions in our latent space by party-wise averaging of MPs' positions, and then fitted a penalized linear regression between these and the party positions in the CHES space. We applied the resulting map to the latent positions of all the MPs and followers, finally obtaining their individual positions along the dimensions of the CHES. In other words, each MP and follower is attributed a continuous position between 0 and 10 on each political dimension measured by the CHES: Left-Right axis, anti-elitism, immigration, EU integration, and so on. For the ease of analysis, in this article we replace values outside of $[0,10]$ by 0 or 10 ; this concerns less than 5\% of the users for each dimension. Detailed statistics are available in the corresponding article \autocite{vendeville2026mapping}. 

For 16 of the most relevant political dimensions, including those that we rely on for the present study, the positions were consistent with self-declared opinions present in the users' bios, demonstrating the validity of the results. Individual indicators of activity and popularity, namely the tweeting rate (mean number of tweets per day) and the number of followers of each user, were extracted from the accounts' metadata given by the API. We lack metadata for 156 users, which represents about 0.015\% of the full dataset and should therefore not impact the results in any significant way. 

We focus the analysis on the followers, as we acknowledge that MPs represent a very specific subgroup of the population, for which a comparison with the general French opinion could be relevant but forms a different research question. 

\subsection{European Social Survey}
The ESS is regularly conducted across European countries. In each country, respondents from a nationally representative panel are asked about their socio-economic situation, religious beliefs, political opinions, and more. We leverage results from edition 4.1 of the 2023 wave \autocite{ESS2023}, in order to compare the opinions of the broad French population with those of our X population. Political opinions are self-declared, given in the form of integers along Likert-type scales of varying ranges.

There are 1,771 respondents in the French panel of ESS in 2023. We discard respondents with at least one missing value in the columns corresponding to the political dimensions that interest us. Missing values include answers ``Not applicable'', ``Refusal'', ``No answer'', ``Don't know''. We end up with 1,417 respondents in the ESS panel.

Finally, note that the ESS contains a question related to online political activity. We show in Supplementary Section 5 that we obtain qualitatively similar results when we only consider respondents who declared being politically active online.

\subsection{Taking into account survey design for ESS data}
Because of the cost-effective way that the ESS is conducted, the data \emph{as is} presents biases. For instance, some individuals and socio-demographic categories are more likely to be surveyed, or to respond, than others. To ensure that our analysis of the ESS panel is robust and representative of the French population, we must take into account the survey design, which consists of three correction variables: weight, clustering, and stratification. We refer the interested reader to the ESS's guide on the matter for more details \autocite{Kaminska2020guide}. 

R's survey package provides a wide range of functions and algorithms that take into account the survey design. We use it to compute the correlation matrices of the ESS panel, to perform PCA for this panel, and to evaluate single-dimensional distributions. For the latter, we create tables with a corrected count of answers at each point of the Likert scale for each dimension. From these tables we draw the histograms, we compute ideological polarization, and we compute Wasserstein distance with the X panel.

\subsection{Binning X panelists}
Along each dimension, ESS panelists have integer-valued opinions, while the opinions of X panelists are continuous in $[0,10]$. To allow comparison, we discretize the X opinion distributions along the same range as ESS, by sorting them into the adequate number of equal-length bins: 11 bins for Left-Right, Immigration, Anti-elitism, EU integration, 5 bins for Social liberalism, Redistribution, Direct democracy, and 6 bins for Environment. For instance, X users with positions on the Left-Right dimension between 0 and $10/11\simeq 0.91$ are sorted in the first bin, users between $0.91$ and $2\times 10/11\simeq 1.82$ in the second bin, and so on. We use these binned distributions for drawing histograms, computing ideological polarization, and computing Wasserstein distance with the ESS panel.

\subsection{Measuring ideological polarization}
There is ongoing scholarly debate about how to measure unidimensional ideological polarization \autocite{dimartino2025quantifying}. We rely on a recently proposed metric \autocite{Bao_Gill_2024}, based on the notion of cumulative entropy. This metric is limited to ordinal data, but has the advantage of simultaneously taking into account spread and bimodality. Let us assume a distribution over the value range $1,...,k$. For instance, the distribution of answers for a survey question with $k$ possible answers. The polarization of the distribution is defined as
\begin{equation}
    E_c = \frac{H(S_1, 1-S_1) + ... + H(S_{k-1}, 1-S_{k-1})}{k-1},
\end{equation}
where $S_j = \sum_{i=1}^j p_i$, and $H$ denotes the binary entropy:
\begin{equation}
    H(p, 1-p) = 2^{-p \log_2 p - (1-p) \log_2 (1-p)}-1.
\end{equation}
We assume that $0\times\log_2 0=0$. The binary entropy $H(p, 1-p) \in [0,1]$ measures the uncertainty of a random variable with a binary outcome. The closer $p$ is to $0.5$, the more equally likely is each possible outcome. Now, $S_j$ is the cumulated mass of the distribution up to point $j$. Thus, $H(S_j, 1-S_j)$ measures the disparity of mass between two halves of the distribution of interest: values smaller or equal to $j$, and values strictly higher. $E_c$ averages this disparity over all possible values $j$. A distribution where all the mass is concentrated at one point will have $E_c=0$, because for each $j$ either $S_j$ or $1-S_j$ is zero, thus $H$ is always zero. By contrast, where the mass is equally split between the two ends of the value range, each $S_j$ is exactly 0.5, maximizing each $H$ term to 1, resulting in $E_c=1$.

\subsection{Wasserstein distance}
The Wasserstein distance ($W_D$), also known as the earth-mover's distance, quantifies the distance between two distributions by the number of unit displacements that are required to transform one distribution into another. It allows us to simply evaluate how far apart are the opinions of the X and the ESS panel. It is defined as 
\begin{equation}
    W_D(p,q) = \inf_{\gamma \in \Gamma(p,q)} \mathbb{E}_{(x,y) \sim \gamma}\vert x-y\vert,
\end{equation}
where $\Gamma(p,q)$ represents the set of all joint distributions with marginals $p$ and $q$. We rescale $W_D$ by its maximum, reached when the masses of $p$ and $q$ are concentrated at the two opposite extremes of the value range. Thus, we obtain values between 0 and 1 for all dimensions, allowing for cross-dimension comparison. We compute $W_D$ using the ready-made implementation in Python's scipy library. 

\subsection{Principal Component Analysis (PCA)}
For dimensionality reduction, we used Principal Component Analysis (PCA). PCA infers a low-dimensional representation of a dataset by leveraging the eigenspace of the correlation matrix. The eigenvectors of the correlation matrix represent pair-wise orthogonal directions of maximum variance in the data. They are the principal components, that we order by the variance of the data they capture, given by the corresponding eigenvalues. Thus, the first principal component is the eigenvector associated with the largest eigenvalue of the correlation matrix, and describes the direction of the original data with the most variance. 

The variance explained (\autoref{fig:expl_var}, panel \textbf{a}) is the ratio of each eigenvalue over the sum of all eigenvalues. Panel \textbf{b} shows a cumulated sum of these ratios, indicating how much variance is explained by the $k$ first PCA components together.

The loadings of the original variables along a PCA component quantify the contribution of each variable to that component. They indicate how strongly and in what direction each variable influences the principal component, helping to interpret what the component represents in terms of the original data. High absolute loading values suggest a strong association between the variable and the component, while low values indicate that the variable is less associated with the component.

\subsubsection{Pre-processing}
Before performing PCA we standardize the datasets, to ensure that variables with high variance do not dominate the analysis simply due to their scale. Standardization transforms all variables to have a mean of 0 and a standard deviation of 1, placing them on a comparable scale. This step is critical because variables with larger original ranges disproportionately influence the principal components. Standardization is even more indispensable for the ESS panels, as the variables span different ranges.

\subsubsection{Rotation of principal components}
Once we have selected the first two principal components for inspection, we rotate them with the varimax method. The purpose of this step is to enhance interpretability of the low-dimensional space. In some cases, it happens that most---if not all---of the variables load on the first components. This makes interpretation difficult, as every dimension is correlated with each component. Varimax rotation consists in finding a different orthogonal basis for the components, which spreads the variance more equally between the components while expressing the same correlation structure. Varimax rotation proceeds by increasing high loadings and decreasing low loadings, effectively enhancing the interpretation of the association between variables and principal components. Note that the ordering between PC1 and PC2 post-rotation is not an indicator of the variance of the original data that each captures. To further enhance readability, we reverse the ordering of each principal component for the ESS panel: we replace coordinates (x,y) in the (PC1,PC2) plane by (-x,-y). This ensures that the ordering of the political dimensions along the rotated components matches that of the X panel.

It is important to note that these steps do not alter the low-dimensional structure of the data. They make it simply more visually interpretable across the two panels. Pre-rotation, pre-reflection loadings are shown in Supplementary Section 2.4.

\subsubsection{Robustness}
We choose to use PCA among other dimensionality reduction methods, in particular because it can take into account the survey design of the ESS, ensuring that our results are representative of the French population. Similar results are obtained with Exploratory Factor Analysis \autocite{Jolliffe2002Principal}, a dimensionality reduction procedure which is widely used for the analysis of political surveys \autocite{stimson2012evolution,bakker2012exploring} (Supplementary Figure 4).

\subsection{Effective dimensionality}
Let $M=8$ be the number of variables in the data, i.e., the number of political dimensions of interest. The effective dimensionality of a panel is defined as
\begin{equation}
    ED = \prod_{m=1}^M (\lambda_m /M)^{-\lambda_m /M},
\end{equation}
where the $\lambda$s are the eigenvalues of the matrix of correlation between the variables. If all the variables are perfectly correlated, there exists a single non-zero eigenvalue equal to $M$ and $ED=1$. If all the variables are perfectly uncorrelated, then there are $M$ eigenvalues equal to $1$ and $ED=M$. Finally, consider the case where the data contains four variables $x_1,x_2,x_3,x_4$. Let us assume that the first two are perfectly correlated and are orthogonal to the latter two, that are also perfectly correlated. The correlation matrix is given by
\begin{equation}
    \begin{pmatrix}
        1 & 1 & 0 & 0 \\
        1 & 1 & 0 & 0 \\
        0 & 0 & 1 & 1 \\
        0 & 0 & 1 & 1
    \end{pmatrix}.
\end{equation}
The eigenvalues are $2,2,0,0$, and we obtain $ED=2$. The data can thus be summarized in a two-dimensional space without loss of information. For instance, each data point is fully described by its $(x_1,x_3)$ coordinates, without needing to know the values of $x_2,x_4$. 

We refer the interested reader to ref.~\autocite{delGiudice2021effective} for a technical introduction to effective dimensionality, and to ref.~\autocite{Koedam_Binding_Steenbergen_2025} for an application to the dimensionality of party positions in E.U. countries.
\section*{Data availability}
The datasets used in this paper are available online at \href{https://osf.io/at5q2/}{https://osf.io/at5q2/} (X dataset), and \href{https://ess.sikt.no/en/}{https://ess.sikt.no/en/} (European Social Survey).

\section*{Code availability}
The code used to analyze data is available online at \href{https://github.com/AntoineVendeville/political-attitudes-socmed-survey}{https://github.com/AntoineVendeville/political-attitudes-socmed-survey}.

\printbibliography

\section*{Acknowledgment}

We thank Alexander T.\ Kindel for his precious advice on the methodology, and Max Falkenberg for useful discussions. This work has been partially funded by the ``European Polarisation Observatory'' (EPO) of CIVICA Research, co-funded by EU's Horizon 2020 programme under grant agreement No 101017201, by European Union Horizon program project ``Social Media for Democracy'' under grant agreement No 101094752 (\url{www.some4dem.eu}), by the \textit{Very Large Research Infrastructure} (TGIR) Huma-Num of CNRS, Aix-Marseille Université and Campus Condorcet, and by Project Liberty Institute project ``AI-Political Machines'' (AIPM).

\section*{Author contribution}

A.V.\ processed the data and performed the analysis. A.V., H.Y.\ and P.R.\ contributed to the design of the study, the interpretation of the results and the writing of the manuscript.

\section*{Competing interests}

The authors declare no competing interests.

\end{multicols}

\end{document}


\title{Political attitudes differ but share a common low-dimensional structure across social media and survey data}

\author[a,b]{Antoine Vendeville}
\author[b,c]{Hiroki Yamashita}
\author[b,a]{Pedro Ramaciotti}

\affil[a]{Sciences Po médialab, Paris, France}
\affil[b]{Complex Systems Institute of Paris Ile-de-France CNRS, Paris, France}
\affil[c]{École des Hautes Études en Sciences Sociales, Paris, France}

\date{}

\maketitle

\tableofcontents

\section{Variables of interest} \label{SI:data}
We detail the variables corresponding to the political dimensions of interest in \autoref{tab:dimensionsSI_x} (X data) and \autoref{tab:dimensionsSI_ess} (ESS data). For the ESS data, we do not list the weight, stratum and primary sampling unit variables (\texttt{anweight, stratum, psu}, respectively). For the X data, we indicate corresponding questions from the 2019 and 2023 waves of the CHES \autocite{ches2019,ches2023}, that were used to calibrate the X data. The scales indicated in the tables are those of the original data. For the purpose of our analysis we reversed the scales of some of them, as indicated in the tables: EU integration (reversed for both X and ESS panels), Anti-elitism, Immigration, Direct democracy (reversed only for ESS panel).

\begin{table*}
\centering
\small
\renewcommand{\arraystretch}{1.5}
\caption{\textbf{Variables of interest in the X dataset.} Variables from the X dataset corresponding to the political dimensions of interest. Scales are continuous between 0 and 10.}
\begin{tabularx}{\textwidth}{|llXX|}
\hline
\textbf{Dimension} & \textbf{Variable} & \textbf{Question (CHES)} & \textbf{Scale} \\
\hline
Left-Right & lrgen\_19 & Position of the party in 2019 in terms of its overall ideological stance. & 0 = Extreme Left, ...., 10 = Extreme Right. \\ \hline
Anti-elitism & antielite\_salience\_23 & How salient has ANTI-ESTABLISHMENT and ANTI-ELITE RHETORIC been to each party? & 0 = Not important at all, ..., 10 = Extremely important \\ \hline
EU integration & eu\_position\_23 & How would you describe the GENERAL POSITION ON EUROPEAN INTEGRATION that the party leadership took over the past three months? & 0 = Strongly opposed, ..., 10 = Strongly in favor \textit{(reversed)}. \\ \hline
Immigration & immigrate\_policy\_19 & Position on immigration policy. & 0 = Strongly favors a liberal policy on immigration, ..., 10 = Strongly favors a restrictive policy on immigration. \\ \hline
Redistribution & lrecon\_23 & Parties can be classified in terms of their stance on ECONOMIC ISSUES such as privatization, taxes, regulation, government spending, and the welfare state. Parties on the economic left want the government to play an active role in the economy. Those on the economic right want a reduced role for government. Where did political parties stand on ECONOMIC issues in the last three months? & 0 = Extreme Left, ..., 10 = Extreme Right. \\ \hline
Direct democracy  & people\_vs\_elite\_1 & Position on people vs elected representatives. Some political parties take the position that ``the people'' should have the final say on the most important issues, for example, by voting directly in referendums. At the opposite pole are political parties that believe that elected representatives should make the most important political decisions. & 0 = Elected oﬃce holders should make the most important decisions, ..., 10 = ``The people'', not politicians, should make the most important decisions. \\ \hline
Environment & environment\_19 & Position towards environmental sustainability. & 0: Strongly supports environmental protection even at the cost of economic growth, ..., 10: Strongly supports economic growth even at the cost of environmental protection. \\ \hline
Social Liberalism & sociallifestyle\_19 & Position on social lifestyle (e.g. rights for homosexuals, gender equality). & 0 = Strongly supports liberal policies, ..., 10 = Strongly opposes liberal policies. \\
\hline
\end{tabularx}
\label{tab:dimensionsSI_x}
\end{table*}

\begin{table*}
\centering
\small
\renewcommand{\arraystretch}{1.5}
\caption{\textbf{Variables of interest in the ESS dataset.} Variables from the ESS dataset corresponding to the political dimensions of interest. Scales are discrete. ``Refusal'', ``Don\'t know'', ``No answer'', ``Not applicable'' count as missing values.}
\begin{tabularx}{\textwidth}{|llXX|}
\hline
\textbf{Dimension} & \textbf{Variable} & \textbf{Question} & \textbf{Answer scale} \\
\hline
Left-Right & lrscale & In politics people sometimes talk of 'left' and 'right'. Where would you place yourself on this scale, where 0 means the left and 10 means the right? & 0: Left, ..., 10: Right, 77: Refusal, 88: Don't know, 99: No answer. \\ \hline
Anti-elitism & trstplt & Please tell me on a score of 0-10 how much you personally trust each of the institutions I read out. 0 means you do not trust an institution at all, and 10 means you have complete trust. Firstly... ...politicians? & 0: No trust at all, ..., 10: Complete trust, 77: Refusal, 88: Don't know, 99: No answer. \textit{(reversed)} \\ \hline
EU integration & euftf & Now thinking about the European Union, some say European unification should go further. Others say it has already gone too far. What number on the scale best describes your position? & 0: Unification already gone too far, ..., 10: Unification go further, 77: Refusal, 88: Don't know, 99: No answer. \textit{(reversed)} \\ \hline
Immigration & imwbcnt & Is France made a worse or a better place to live by people coming to live here from other countries? & 0: Worse place to live, ..., 10: Better place to live,  77: Refusal, 88: Don't know, 99: No answer. \textit{(reversed)} \\ \hline
Redistribution & gincdif & Please say to what extent you agree or disagree with each of the following statements. The government should take measures to reduce differences in income levels. & 1: Agree strongly, 2: Agree, 3: Neither agree nor disagree, 4: Disagree, 5: Disagree strongly, 7: Refusal, 8: Don't know, 9: No answer. \\ \hline
Direct democracy  & psppsgva & How much would you say the political system in France allows people like you to have a say in what the government does? & 1: Not at all, 2: Very little, 3: Some, 4: A lot, 5: A great deal, 7: Refusal, 8: Don't know, 9: No answer. \textit{(reversed)} \\ \hline
Environment & impenva & Now I will briefly describe some people. Please listen to each description and tell me how much each person is or is not like you. She/he strongly believes that people should care for nature. Looking after the environment is important to her/him. & 1: Very much like me, 2: Like me, 3: Somewhat like me, 4: A little like me, 5: Not like me, 6: Not like me at all, 66: Not applicable, 77: Refusal, 88: Don't know, 99: No answer. \\ \hline
Social Liberalism & freehms & Please say to what extent you agree or disagree with each of the following statements. Gay men and lesbians should be free to live their own life as they wish. & 1: Agree strongly, 2: Agree, 3: Neither agree nor disagree, 4: Disagree, 5: Disagree strongly, 7: Refusal, 8: Don't know, 9: No answer. \\
\hline
\end{tabularx}
\label{tab:dimensionsSI_ess}
\end{table*}

\section{Low-dimensional structure} \label{SI:reduced_dim}
We show additional results concerning the low-dimensional structure of the opinion spaces.


\subsection{Contributions of dimensions to principal components} \label{SI:contributions}
\begin{figure*}[t]
    \centering
    \includegraphics[width=\textwidth]{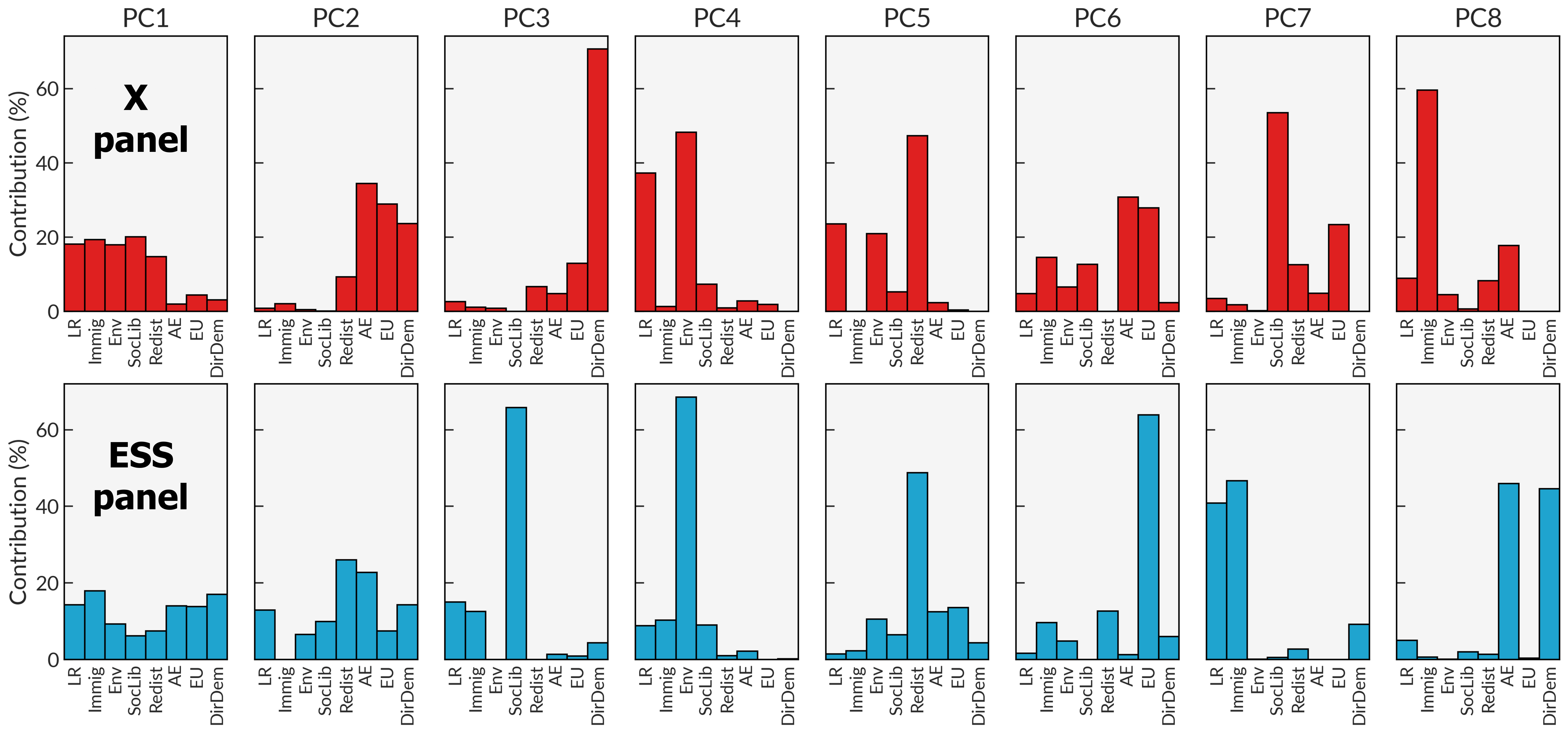}
    \caption{\textbf{Contribution of dimensions to the principal components.} For each principal component we first square the loadings and then divide the results by the sum of those squares. These are computed before rotation and reflection of the principal components.}
    \label{fig:contributions}
\end{figure*}

\autoref{fig:contributions} shows the contributions (normalized squared loadings) of each dimension to each principal component. Because rotation is dependent on the number of rotated components, rotating components beyond the first two that we analyze in the article would output inconsistent results with the rest of our analysis. Thus we show here loadings along the unrotated principal components---note that they are mostly the same as the rotated components for the X panel (\autoref{fig:shining_suns_unrot}). The composition of the third principal components consists almost exclusively of Direct democracy (X panel) or Social liberalism (ESS panel). This echoes observations regarding the loadings of these dimensions on the first two principal components in the main text.

\subsection{Distribution of users along principal components} \label{SI:pc_distrib}
\begin{figure*}
    \centering
    \includegraphics[width=\textwidth]{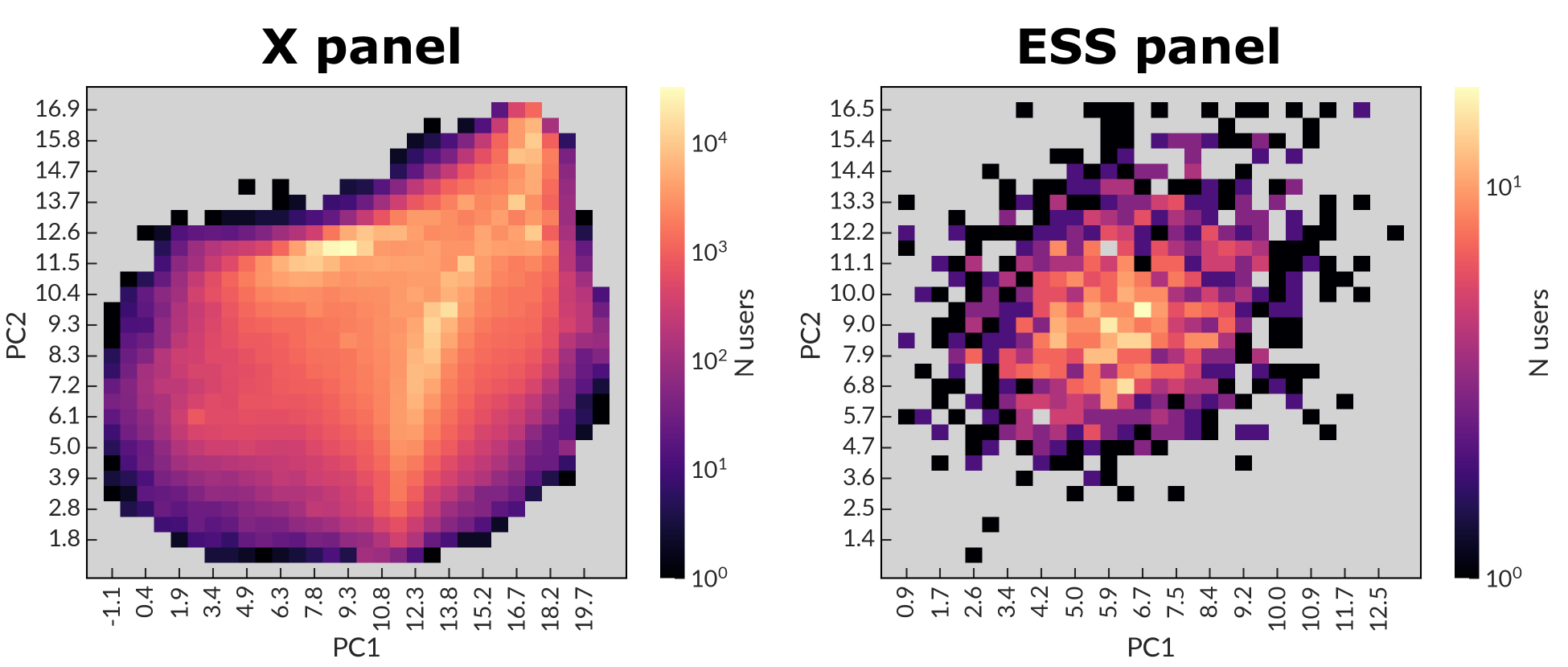}\caption{\textbf{Distribution of users in the PC space.} The (PC1,PC2) plane is divided in a $30\times 30$ grid. Cells are colored according to the log number of users they contain.}
    \label{fig:user_density}
\end{figure*}

\begin{figure*}
    \centering
    \includegraphics[width=.95\textwidth]{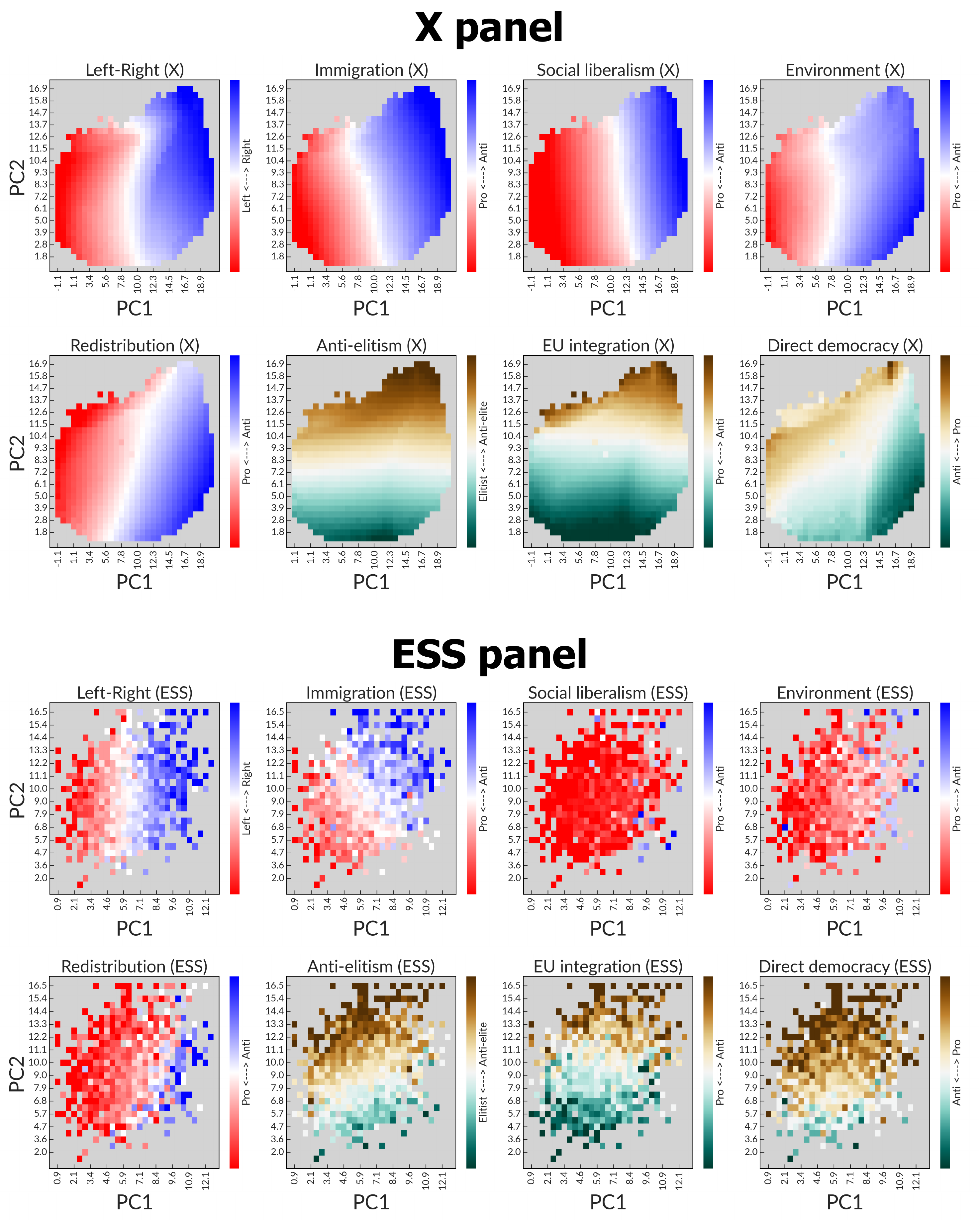}
    \caption{\textbf{Opinion distributions along PC1 and PC2.}  The (PC1,PC2) plane is divided in a $30\times 30$ grid. Cells are colored according to the average position of the panelists on the corresponding dimension.}
    \label{fig:maps}
\end{figure*}

\autoref{fig:user_density} shows the log density of users in the two-dimensional (PC1,PC2) plane, divided into a $30\times 30$ grid. \autoref{fig:maps} shows, for each political dimension, the average opinion of users in the (PC1,PC2) plane.

\subsection{Exploratory factor analysis} \label{SI:efa}
To confirm the robustness of the results we obtained via PCA, we also perform an exploratory factor analysis (EFA). EFA differs from PCA as it explicitly posits a latent model to express the data in a low-dimensional space. The assumption is that each variable in the data can be expressed as a linear combination of factors, plus an error term. This means that the number of factors must be chosen a priori, as the results on early factors change when adding new factors---contrary to PCA. Additionally, EFA focuses on reproducing the covariance between the variables, while PCA focuses on capturing the variance of individual variables. We refer the interested reader to ref.~\autocite{Jolliffe2002Principal} (Chapter 7) for an introduction to EFA.

\begin{figure*}
    \centering
    \includegraphics[width=\textwidth]{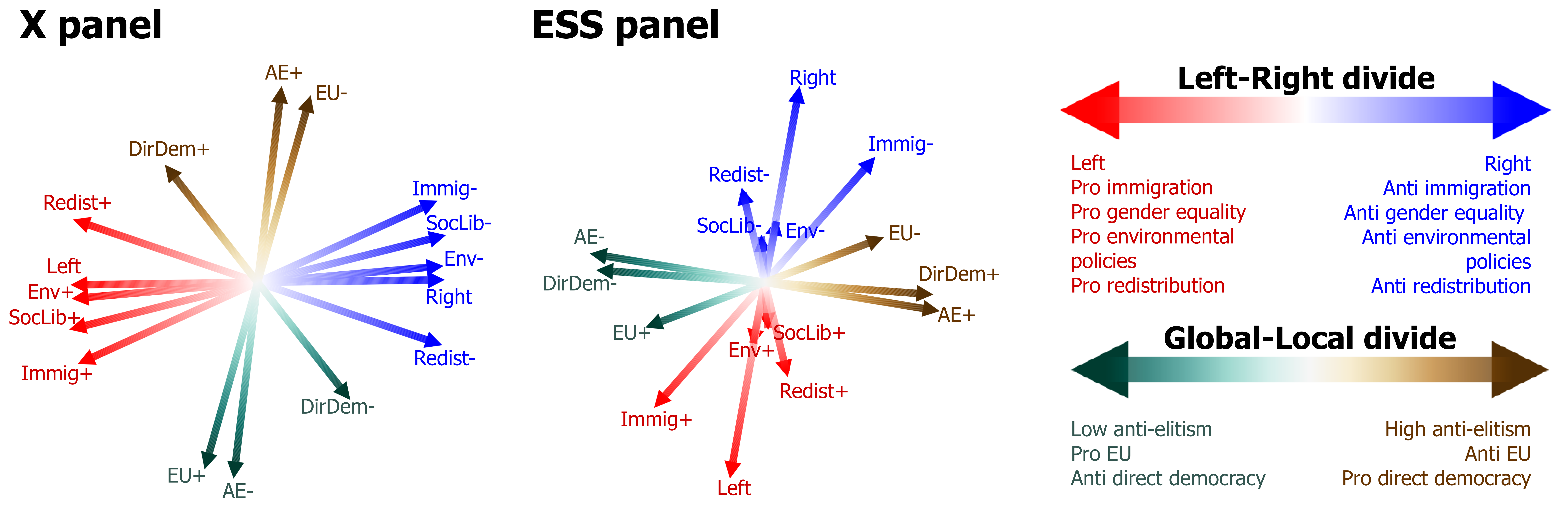}
    \caption{\textbf{Loadings of political dimensions obtained via exploratory factor analysis.} Each arrow shows the loading of a political dimension along the first two factors. Arrow length is proportional to the absolute value of the loading. Red-blue arrows correspond to dimensions related to the left-right attitudes, and green-brown arrows correspond to dimensions related to global-local attitudes Plus (+) signs are short for ``pro'', and minus (-) signs for ``anti''.}
    \label{fig:EFA}
\end{figure*}

We set the number of factors to two. As with PCA, we apply a varimax rotation\footnote{Note that rotating the factors is standard practice for EFA.}, and reflect the factors for enhanced interpretability. \autoref{fig:EFA} shows the loadings of the political dimensions along the factors. We obtain similar structures as with PCA, although we note the greater variability in the absolute value of loadings for the ESS panel.

\subsection{Unrotated principal components} \label{SI:unrotated}
\begin{figure*}
    \centering
    \includegraphics[width=\textwidth]{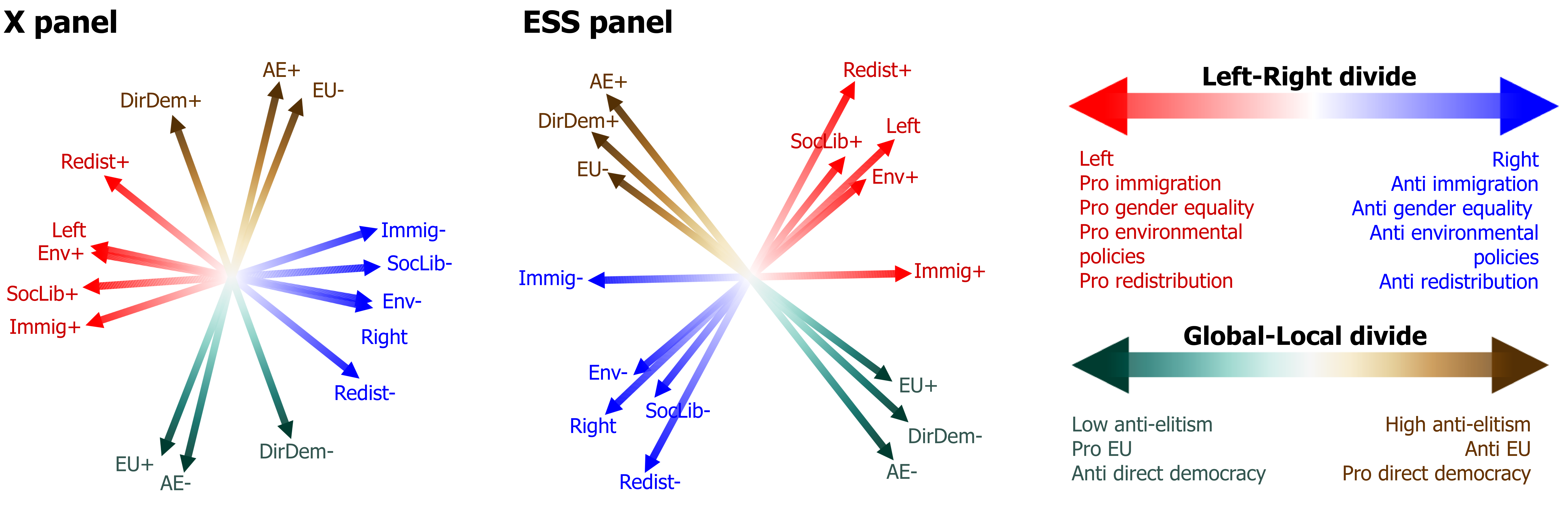}
    \caption{\textbf{Loadings of political dimensions along principal components without rotation or reflectino.} Each arrow shows the loading of a political dimension along the first two principal components, PC1 (x-axis) and PC2 (y-axis). Arrow length is proportional to the absolute value of the loading. Red-blue arrows correspond to dimensions related to the left-right attitudes, and green-brown arrows correspond to dimensions related to global-local attitudes Plus (+) signs are short for ``pro'', and minus (-) signs for ``anti''.}
    \label{fig:shining_suns_unrot}
\end{figure*}

\begin{figure*}
    \centering
    \includegraphics[width=.9\textwidth]{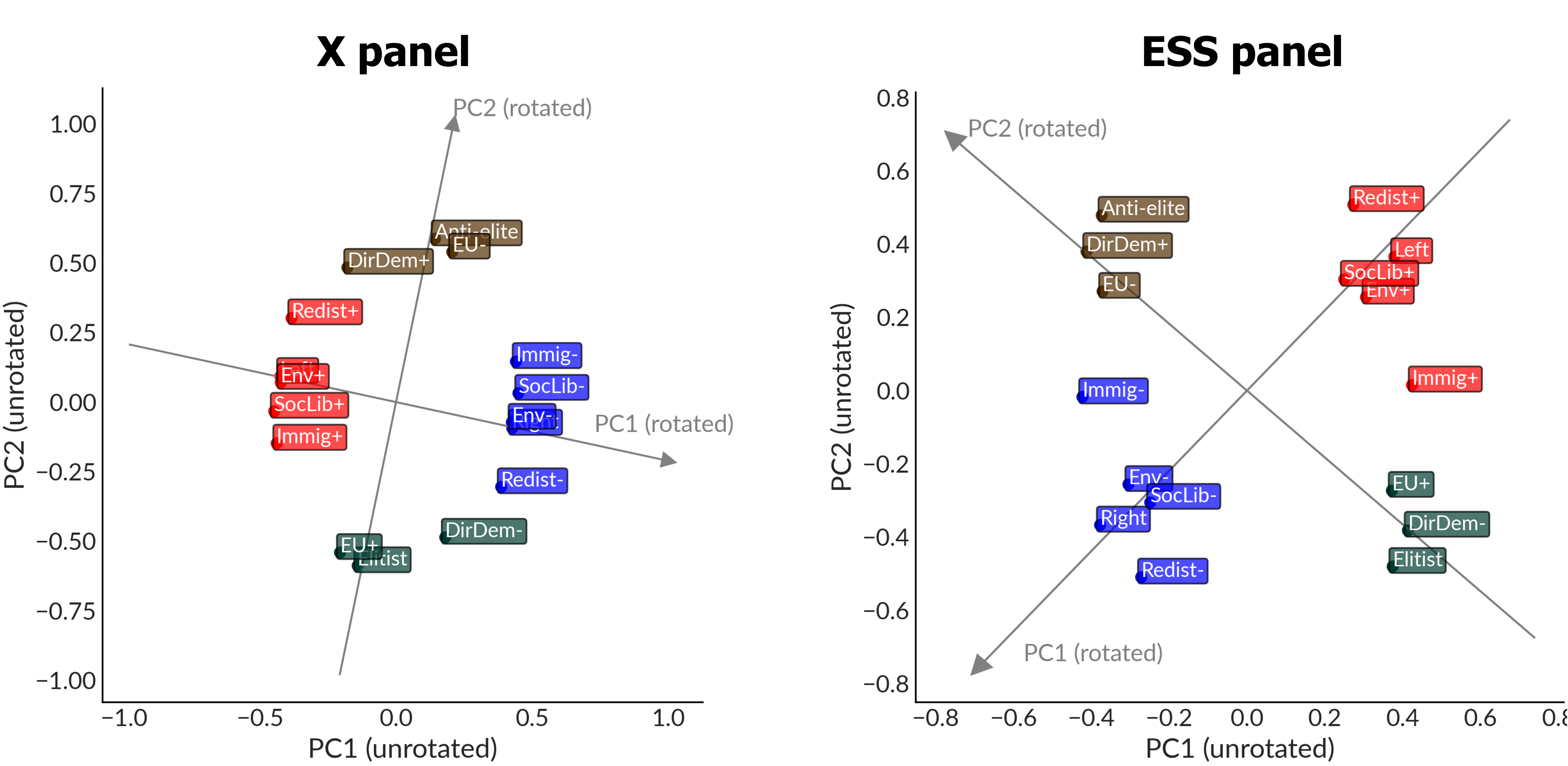}
    \caption{\textbf{Rotated versus unrotated principal components.} The x and y-axis delineate the unrotated principal components. Boxes' coordinates correspond to the loadings of the dimensions, symmetrized around the origin. Arrows show the coordinates of the rotated principal components.}
    \label{fig:rot_vs_unrot}
\end{figure*}

\autoref{fig:shining_suns_unrot} shows the loadings of the first two principal components before applying varimax rotation and before applying reflection. \autoref{fig:rot_vs_unrot} shows how the principal components are rotated by varimax, by plotting the rotated axes in the unrotated PCA space, alongside the (unrotated) loadings of dimensions.

\section{Distribution of activity, popularity and visibility} \label{SI:actipopvis_distrib}
\begin{figure*}[t]
    \centering
    \includegraphics[width=.9\textwidth]{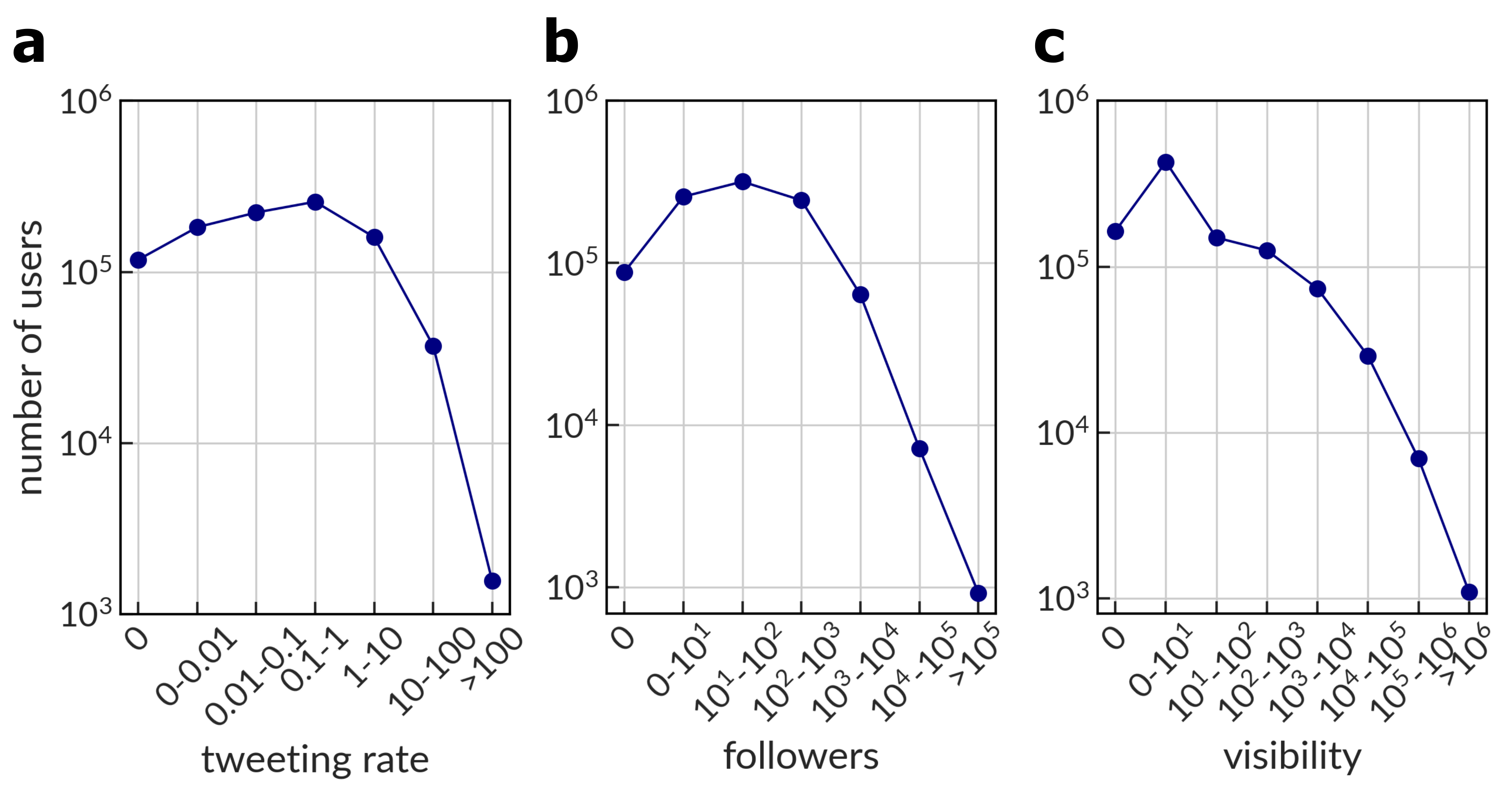}
    \caption{Number of users in each activity (a), popularity (b), and visibility (c) bin.}
    \label{fig:n_users_activity_popularity}
\end{figure*}

\autoref{fig:n_users_activity_popularity} shows the number of users in each activity, popularity, and visibility bin, in log scale.

\section{Visibility analysis} \label{SI:visibility}
This section concerns the analysis of the X panel, where panelists are weighted by their estimated visibility on the platform. We first provide additional results regarding the distribution of positions and the low-dimensional structure weighted X panel. Then, we justify our estimation of visibility by comparing it with impression counts received by the tweets of a subsample of the users.

\subsection{Dimension-wise distributions of positions}
We show the weighted distribution of X panelists along the eight political dimensions of interest, compared with the unweighted panel. The results show distributions closer to stances on all dimensions.

\begin{figure*}
    \centering
    \includegraphics[width=\textwidth]{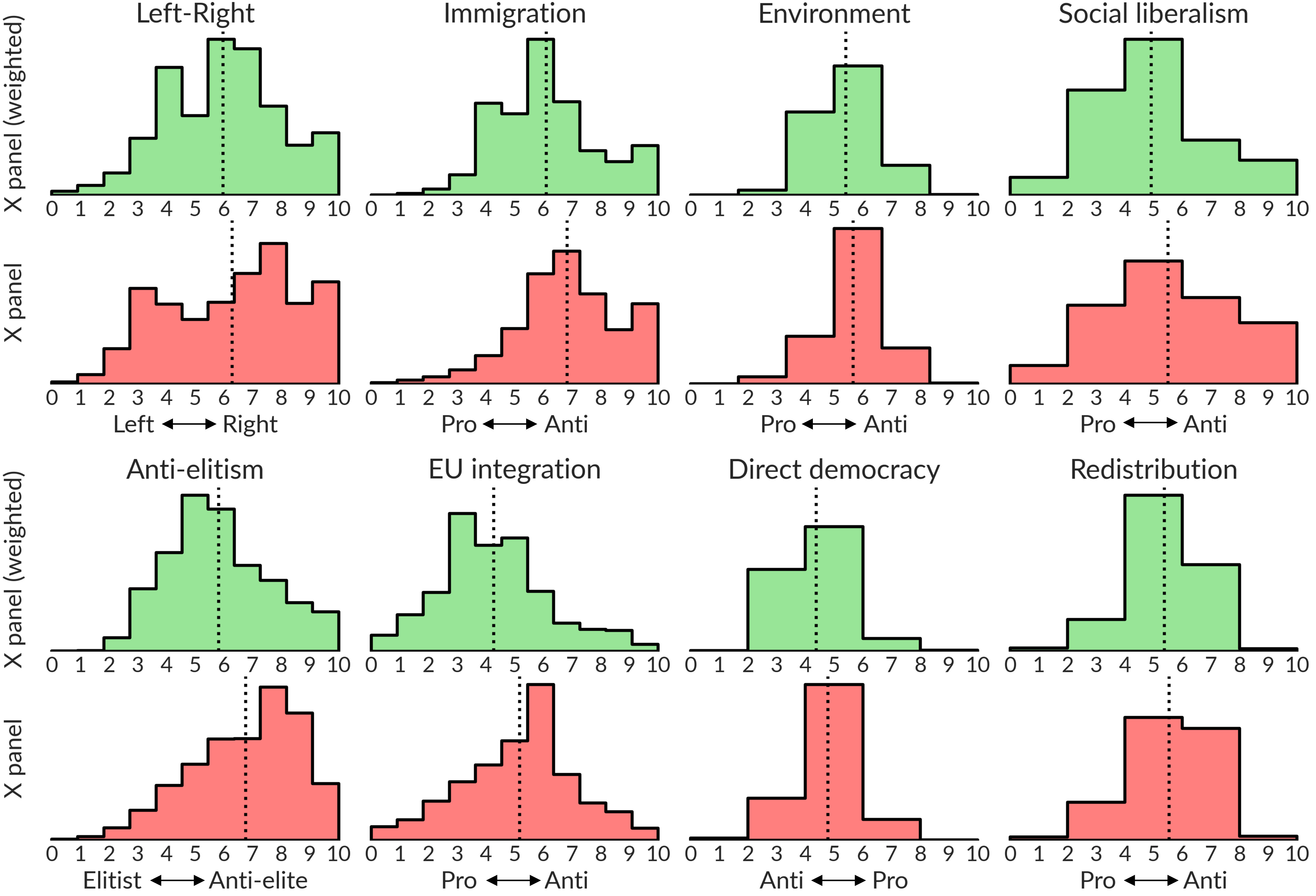}
    \caption{\textbf{Opinion distributions of X panel, weighted by visibility.} Dotted vertical lines indicate the mean. Green: weighted X panel, Red: unweighted X panel.}
    \label{fig:histograms_visibility}
\end{figure*}

\subsection{Low-dimensional structure}
\begin{figure*}
    \centering
    \includegraphics[width=\textwidth]{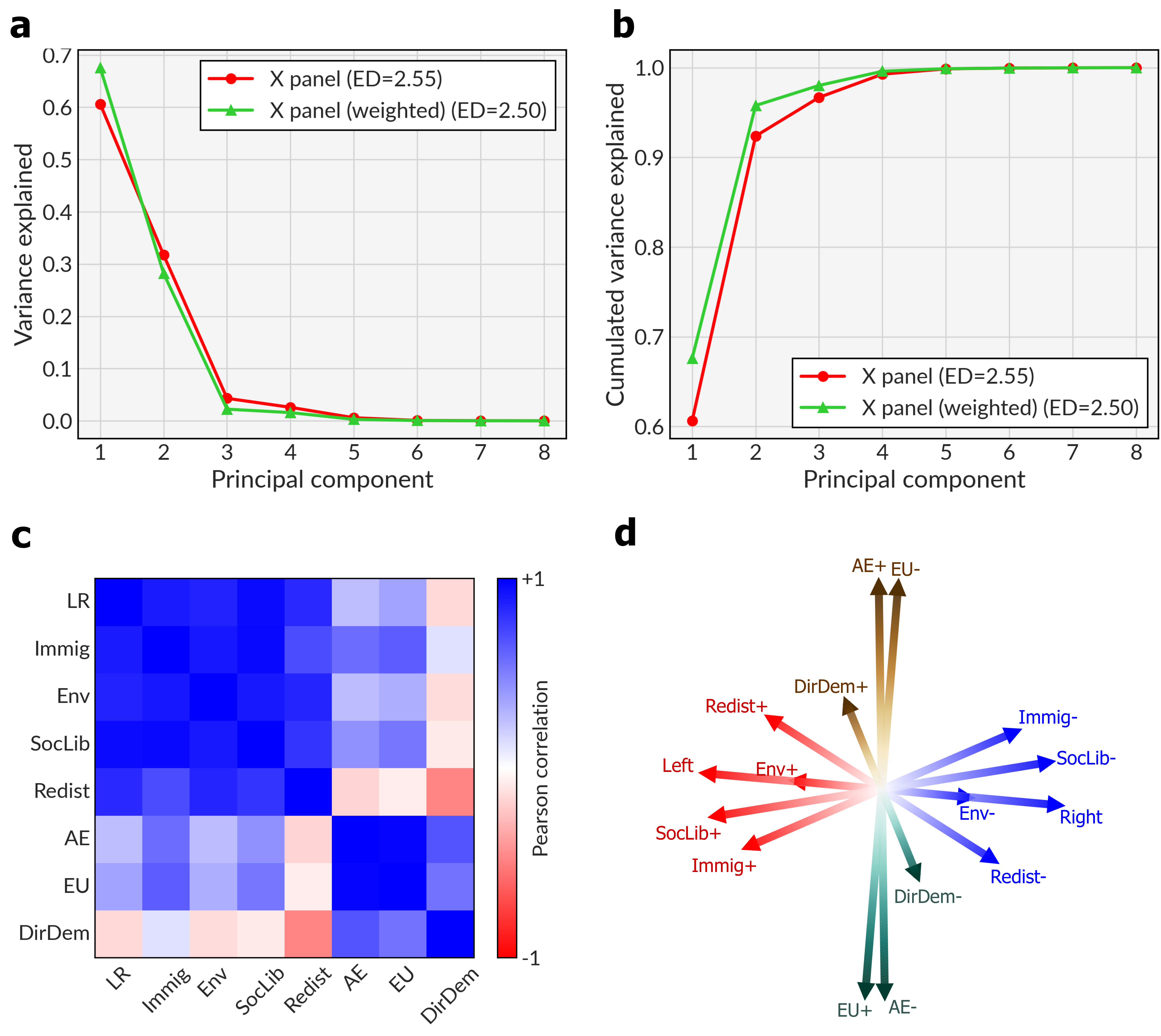}
    \caption{\textbf{Low-dimensional structure of the X panel, weighted by visibility.} \textbf{(a)} Variance explained by principal components. \textbf{(b)} Cumulated variance explained by principal components. \textbf{(c)} Weighted Pearson correlation matrix. All correlations are significant at the level $p<0.001$. \textbf{(d)} Loadings of the political dimensions along the first two principal components.}
    \label{fig:lowdim_visibility}
\end{figure*}

When weighting X panelists by their visibility, the low-dimensional structure of political positions does not exhibit important changes. To identify this structure, we first compute a weighted correlation matrix. We then apply PCA on this correlation matrix, performing rotation of reflection of components afterwards to enhance interpretability (see ``Methods'' in the main text). We show the low-dimensional structure of the weighted X panel in \autoref{fig:lowdim_visibility}.

\subsection{Variations along visibility values}
\begin{figure*}
    \centering
    \includegraphics[width=\textwidth]{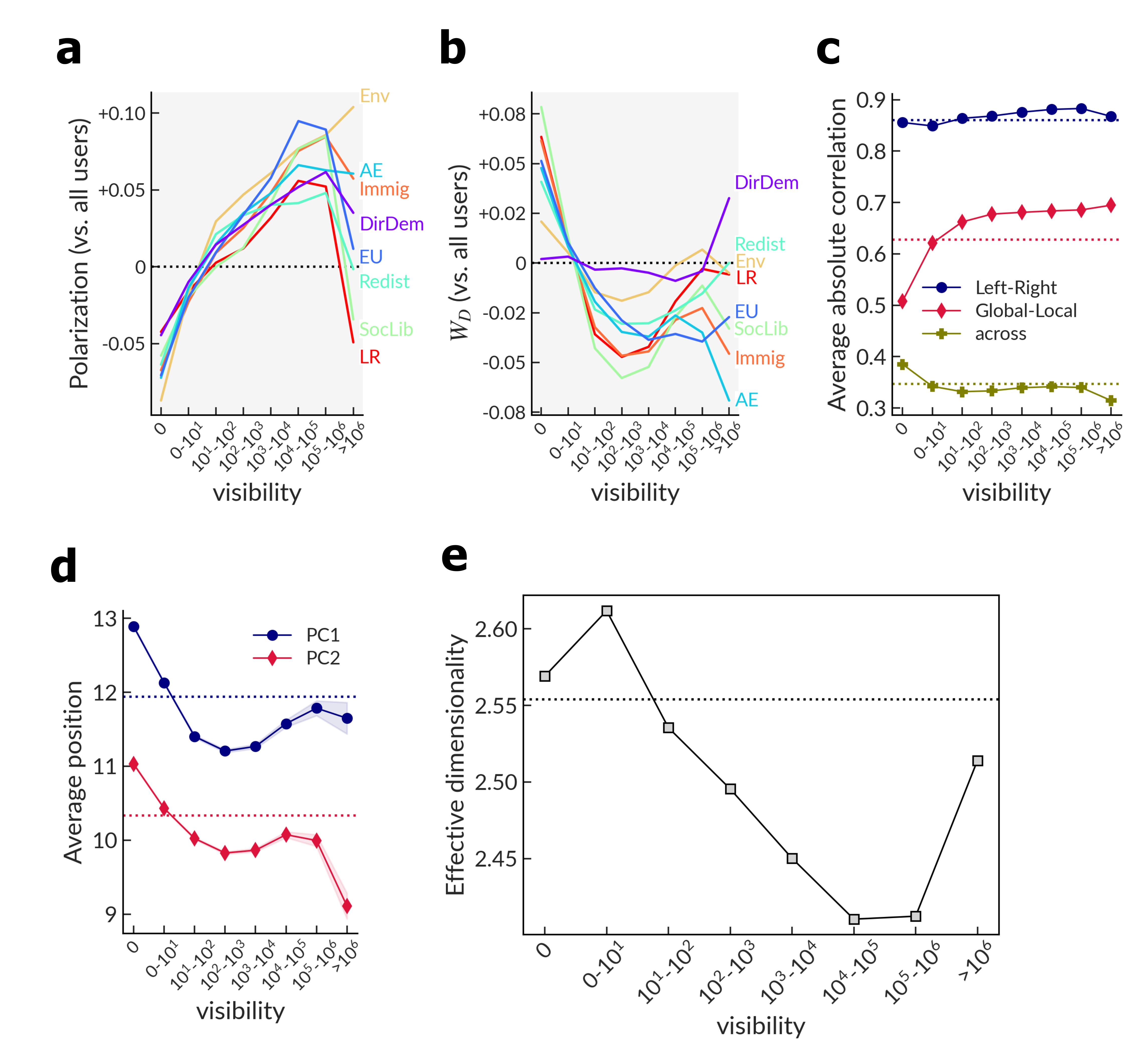}
        \caption{\textbf{Variation of polarization, representativeness and dimensionality of X panelists with visibility.} Users are binned according to their estimated visibility ($tweeting\_rate \times followers$). Dotted lines indicate values computed for the whole panel. \textbf{(a)} Polarization along the different dimensions: in-bin value minus overall value. \textbf{(b)} Wasserstein distance between positions of X users and ESS respondents: in-bin value minus overall value. \textbf{(c)} Average absolute correlation between dimensions within the Left-Right divide (blue circles), within the Global-Local divide (red diamonds), and across the Left-Right and Global-Local divide (green crosses). \textbf{(d)} Average position of users along PC1 (blue circles) and PC2 (red diamonds), with 95\% confidence intervals. \textbf{(e)} Effective dimensionality.}
    \label{fig:visibility}
\end{figure*}

\autoref{fig:visibility} shows how our metrics of interest (polarization, Wasserstein distance, dimensionality, PC1 and PC2 distributions) evolve with visibility. The most visible users exhibit opinion distributions closer to those of the ESS panel (subplot b). The more visible they are, the more left-wing and globalist they get. The only surprising result is the lower effective dimensionality, while the explained variance of PC1 shrinks and that of PC2 grows with visibility.

\subsection{Robustness of visibility metric}
To justify our choice of visibility metric, we leverage a database of tweets collected during the whole duration of January and February 2023, obtained by querying all tweets in French containing a URL. That includes tweets containing an image or video. When the tweet was a reply, the whole thread and original tweet were also collected. This database totals about 7.9M tweets, covering 97,947 users of the X panel (10\%). In particular, this database contains the \emph{impression count} of each tweets\footnote{As per the API: ``A count of how many times the Post has been viewed (not unique by user). A view is counted if any part of the Post is visible on the screen.'' (\url{https://docs.x.com/x-api/fundamentals/metrics}).}. 

\begin{figure*}[t]
    \centering
    \includegraphics[width=.5\textwidth]{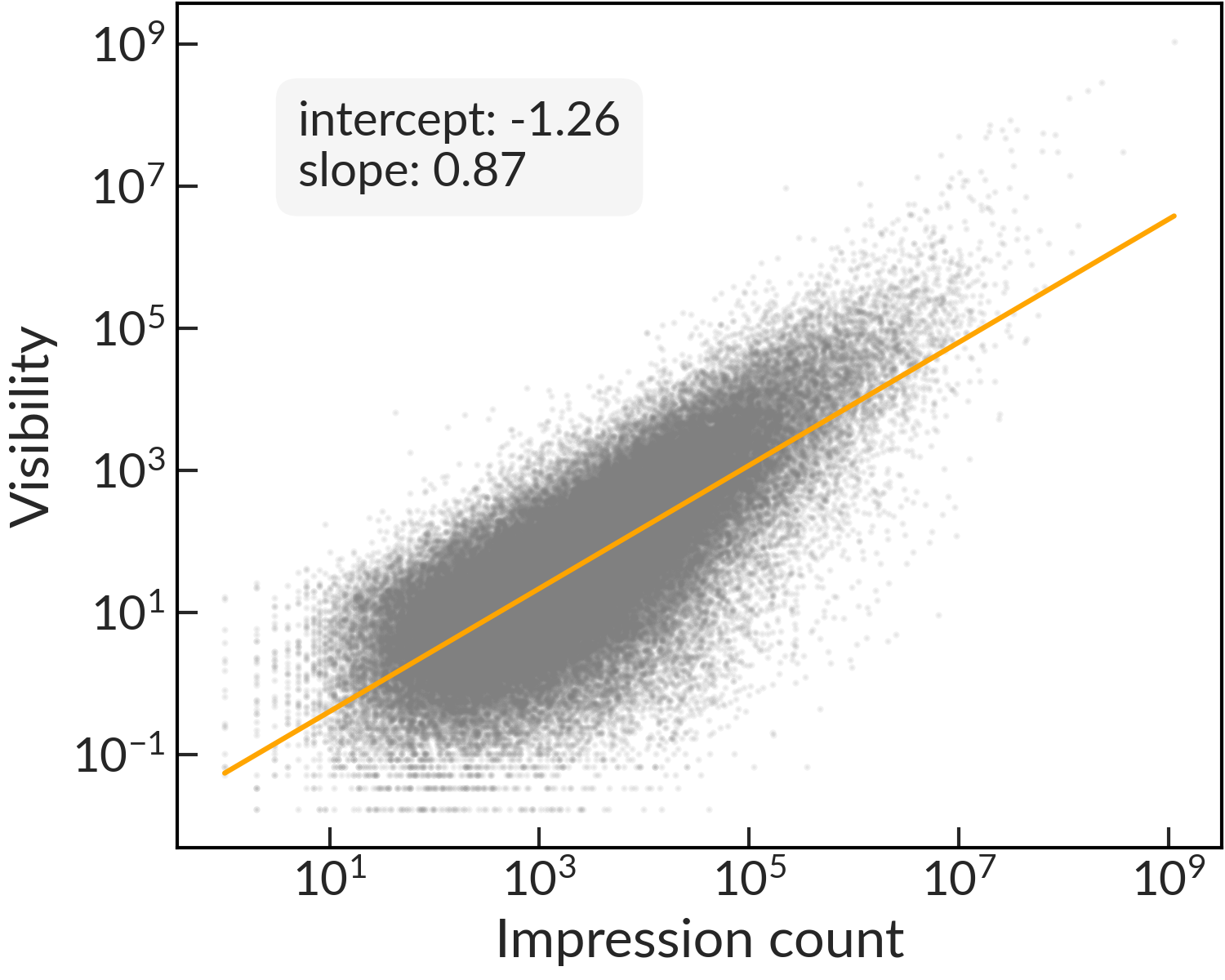}
    \caption{Regression plot between impression count (x-axis) and visibility (y-axis). Grey dots indicate individual values.}
    \label{fig:visibility_vs_ic_regplot}
\end{figure*}

We compute the total impression count for the users of our X panel that are present in this database. We compare these counts with our visibility metric, defined by $tweeting\_rate \times followers$. The tweeting rate is computed on the basis of the number of tweets for which we have impression counts, divided by the number of days spanned by our data (31 in January plus 28 in February). We find a Pearson correlation of 0.92 $(p<0.001)$ and a Spearman correlation of 0.74  $(p<0.001)$. We regress visibility against impression counts in \autoref{fig:visibility_vs_ic_regplot}. Because values span several orders of magnitude, we must employ a log scale. Thus, we discard users with visibility=0 (final user count: 97,848).

\begin{figure*}
    \centering
    \includegraphics[width=\textwidth]{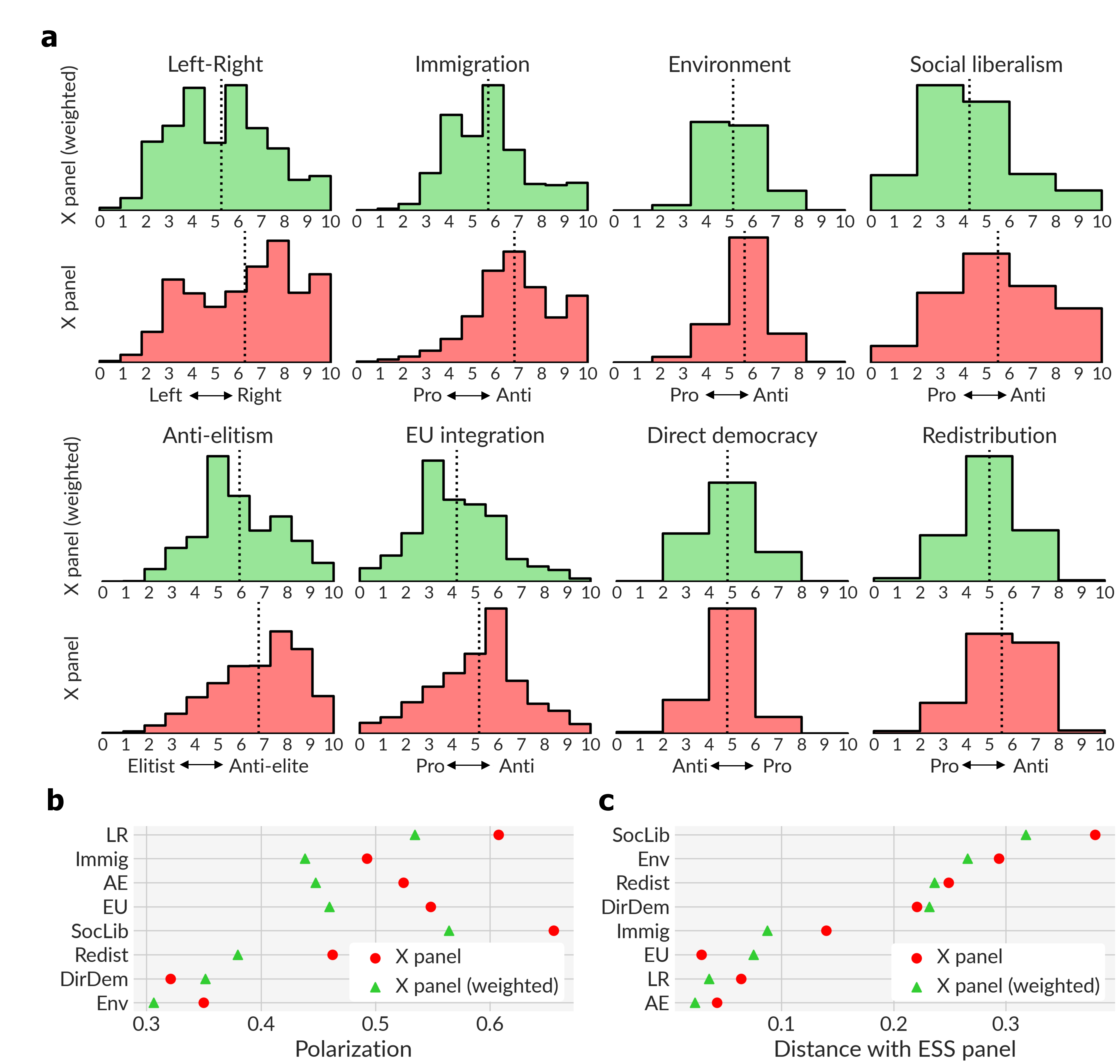}
    \caption{\textbf{Opinion distributions of X panel, weighted by impression counts.} Dotted vertical lines indicate the mean. Green: weighted X panel, Red: unweighted X panel.}
    \label{fig:histograms_ic}
\end{figure*}

\begin{figure*}
    \centering
    \includegraphics[width=\textwidth]{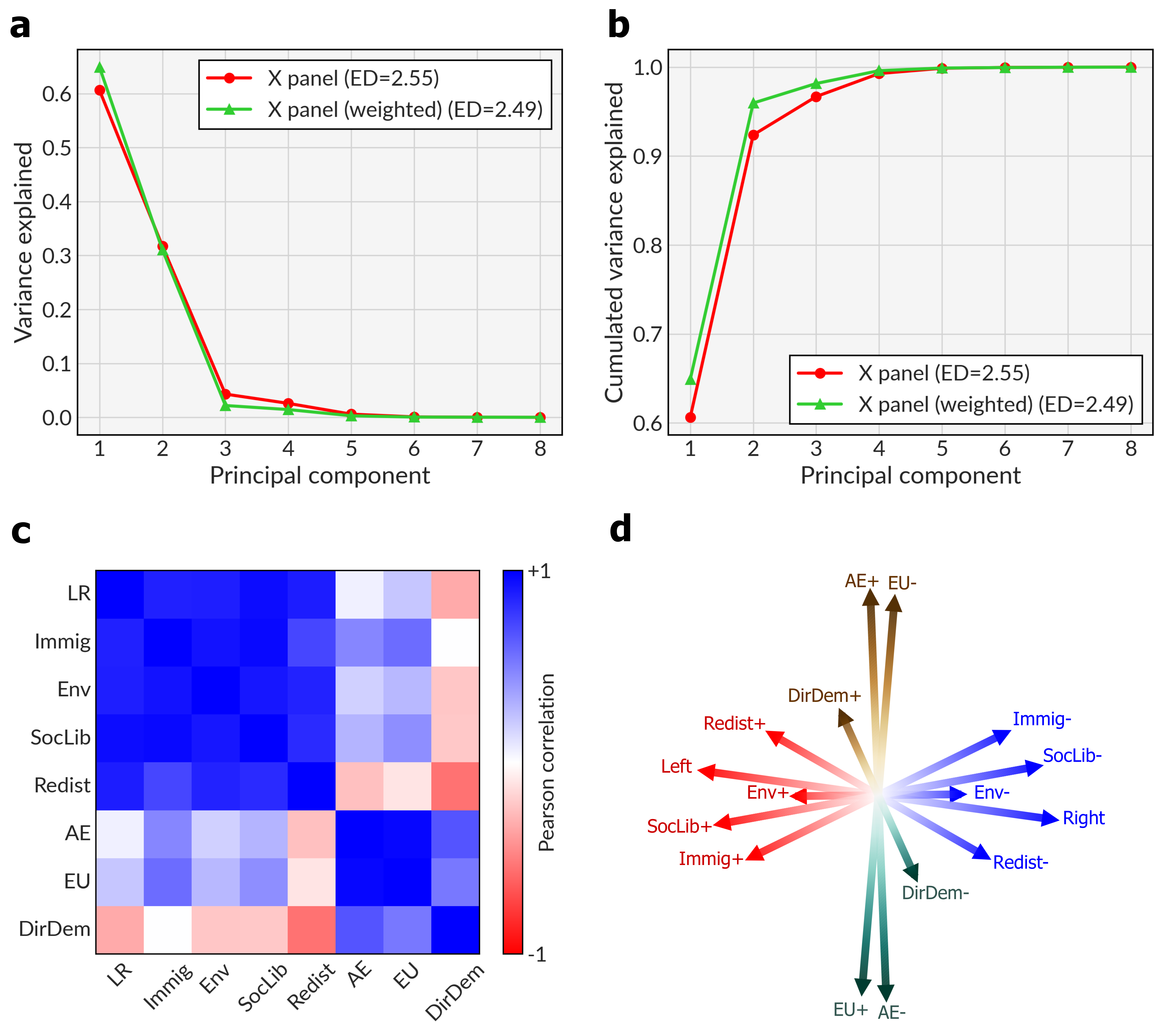}
    \caption{\textbf{Low-dimensional structure of the X panel, weighted by impression count.} \textbf{(a)} Variance explained by principal components. \textbf{(b)} Cumulated variance explained by principal components. \textbf{(c)} Weighted correlation matrix. All values are significant at the level $p<0.001$. \textbf{(d)} Loadings of the political dimensions along the first two principal components.}
    \label{fig:lowdim_ic}
\end{figure*}

In addition, we analyze unidimensional distributions of positions (\autoref{fig:histograms_ic}) and low-dimensional structure of the X panel, weighted by impression counts (\autoref{fig:lowdim_ic}). The analysis is thus restricted to the 98K users for which we have access to this metric.  Results are highly similar to those obtained by using our visibility metric, indicating that it is a good proxy for impression count.

\section{Politically active online ESS respondents} \label{SI:X_vs_ESS-PAO}
We derive a secondary dataset called \emph{ESS-PAO} (``politically active online''), by restricting ourselves to ESS respondents who declared having posted online content related to politics (variable name: \texttt{pstplonl}). We expect this subgroup to exhibit opinions that are closer to those of our X population. 

\begin{figure*}
    \centering
    \includegraphics[width=\textwidth]{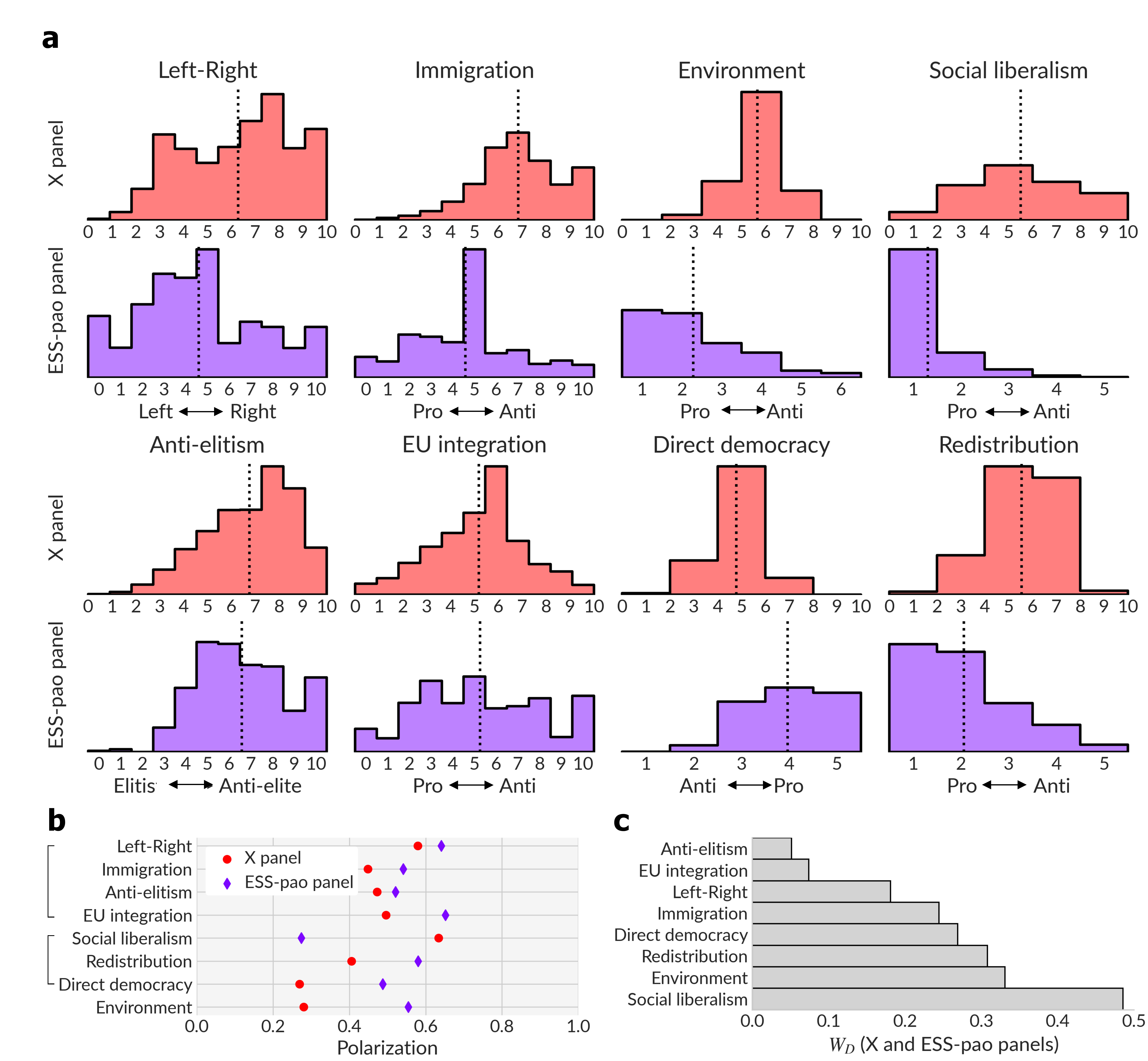}
    \caption{\textbf{Comparison between opinions of X users and ESS-PAO respondents.} (a) Histograms of opinion distributions. Dashed vertical lines indicate averages. Colored half-panes in the background mark the side on which the distributions are concentrated. Distributions for the X panel are sorted into bins to facilitate comparison with the ESS panel. (b) Polarization values across dimensions. Brackets indicate groups of dimensions that share the same support, and within which polarization values can be compared. (c) Normalized Wasserstein distance between X and ESS-PAO panelists (purple diamonds), and between X and ESS panelists (blue squares).}
    \label{fig:epo_vs_ess_opa}
\end{figure*}

\autoref{fig:epo_vs_ess_opa} shows dimension-wise opinion distributions of X panelists and ESS-PAO panelists. Opinion distribution of X panelists are discretized by sorting them into bins, to facilitate comparison with ESS-PAO panelists. Results are similar as for the whole ESS panel, with two noticeable differences. Distributions of opinions for ESS-PAO panelists does not reflect the predominantly right-wing positioning of X users, but better reflects their anti-EU sentiment. Polarization levels in ESS-PAO are slightly higher than in ESS. Surprisingly, the $W_D$ values are also (slightly) higher for most dimensions, meaning that unidimensional opinion distributions of the X panel are further away from the subsample of ESS respondents that are politically active online, although we note that more than $10,000$ of the $900,000$ X panelists do not post (\autoref{fig:n_users_activity_popularity}).

\begin{figure*}
    \centering
    \includegraphics[width=\textwidth]{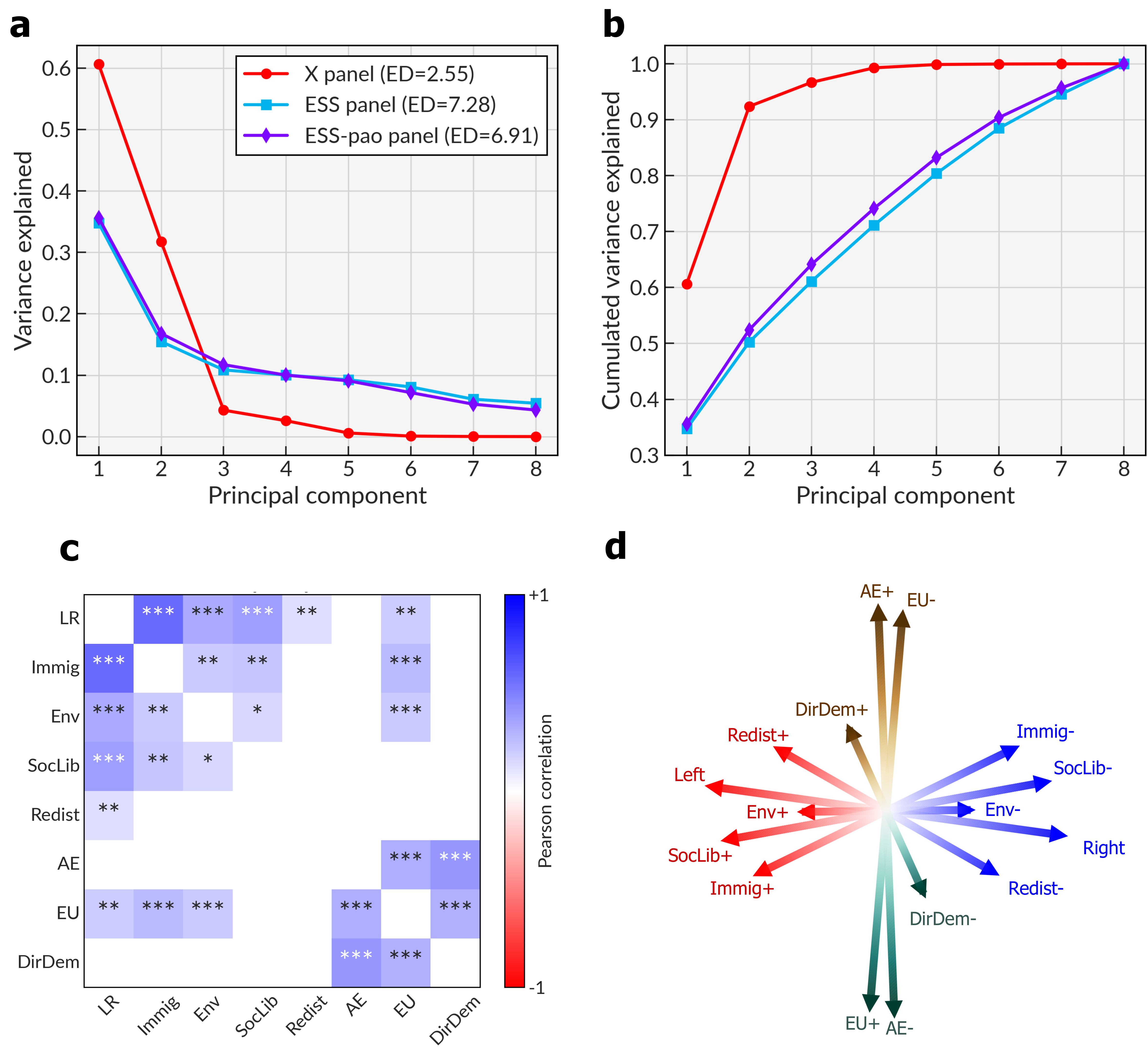}
    \caption{\textbf{Low-dimensional structure of opinions for ESS-PAO respondents.} (a) Explained variance by principal component and effective dimensionality. (b) Cumulated explained variance. (c) Correlation matrix. Percent of significant correlations: 46\%, average value of significant off-diagonal correlations: 0.29, strongest off-diagonal correlation: $0.58$ (Left-Right and immigration). (d) Loadings of political dimensions along principal components, rotated and reflected to enhance interpretability.}
    \label{fig:lowdim_opa}
\end{figure*}

\autoref{fig:lowdim_opa} illustrates the low-dimensional structure of political positions for ESS-PAO panelists. The ESS-PAO panel exhibits stronger correlations than the full ESS panel but fewer are significant, with very few significant correlations between Left-Right and Global-Local attitudes, restricted to EU integration. This leads us to believe that the hypothesized Left-Right and Global-Local divides appear more salient among ESS-PAO respondents than for the whole ESS panel. ESS-PAO exhibits a slightly lower dimensionality than ESS, with six components being sufficient to explain 90\% of the variance, meaning that politically active online citizens have slightly more constrained opinions. The overall configuration of dimensions, given by the loadings along principal components, is similar to that of ESS panelists.

\printbibliography